\documentclass[aps, prb,
reprint,
superscriptaddress,
longbibliography]{revtex4-2}

\usepackage{amsmath}
\usepackage{amssymb}
\usepackage{amsfonts}
\usepackage{physics}
\usepackage{enumitem}
\setlist{listparindent = \parindent}

\usepackage[cmyk]{xcolor}
\definecolor{db}{cmyk}{1, 1, 0, 0.25}
\definecolor{dr}{cmyk}{0, 0, 0, 1}

\usepackage{hyperref}
\hypersetup{colorlinks, linkcolor = blue, urlcolor = blue, citecolor = blue}
\usepackage{graphicx}

\renewcommand{\t}{\text}

\begin{document}

\title{Enhanced Thermalization of Exciton-Polaritons in Optically Generated Potentials}

\author{Yoseob Yoon}
\email{yoonys@alum.mit.edu}
\thanks{Present address: Department of Physics, University of California, Berkeley, CA 94720, USA and Materials Sciences Division, Lawrence Berkeley National Laboratory, Berkeley, CA 94720, USA.}
\affiliation{Department of Chemistry, Massachusetts Institute of Technology, Cambridge, MA 02139, USA}

\author{Jude Deschamps}
\affiliation{Department of Chemistry, Massachusetts Institute of Technology, Cambridge, MA 02139, USA}

\author{Mark Steger}
\affiliation{National Renewable Energy Laboratory, Golden, CO, 80401, USA}

\author{Ken W.\ West}
\author{Loren N.\ Pfeiffer}
\affiliation{Department of Electrical Engineering, Princeton University, Princeton, NJ 08544, USA}

\author{David W.\ Snoke}
\affiliation{Department of Physics, University of Pittsburgh, Pittsburgh, PA 15260, USA}

\author{Keith A.\ Nelson}
\email{kanelson@mit.edu}
\affiliation{Department of Chemistry, Massachusetts Institute of Technology, Cambridge, MA 02139, USA}

\date{\today}

\begin{abstract}
Equilibrium Bose-Einstein condensation of exciton-polaritons, demonstrated with a long-lifetime microcavity [\href{https://doi.org/10.1103/PhysRevLett.118.016602}{Phys.\ Rev.\ Lett.\ \textbf{118}, 016602 (2017)}], has proven that driven-dissipative systems can undergo thermodynamic phase transitions in the limit where the quasiparticle lifetime exceeds the thermalization time. Here, we identify the role of dimensionality and polariton interactions in determining the degree of thermalization in optically generated traps. To distinguish the effect of trapping from interactions and lifetimes, we measured the polariton distribution under four nonresonant Gaussian pumps in a square geometry and compared it with polariton distributions measured with each pump individually. We found that significant redistribution of polaritons arises by trapping and modification of the density of states. Surprisingly efficient polariton-polariton scattering below the condensation threshold is evidenced by the depletion of the inflection-point polaritons. Our work provides a deeper understanding of polariton distributions and their interactions under various geometries of optically generated potentials.
\end{abstract}

\maketitle

\section{Introduction}
Hybrid light-matter quasiparticles in a quantum-well microcavity, called exciton-polaritons (hereafter ``polaritons''), arise from the collective strong coupling between excitons and cavity photons \cite{Yamamoto1999, Deng2010, Carusotto2013}. Even with equal mixtures of excitons and photons, polaritons have \emph{photon-like} effective mass, allowing macroscopic quantum phenomena such as Bose-Einstein condensation (BEC) up to room temperature \cite{Deng2002, Kasprzak2006, Balili2007, Christopoulos2007, Plumhof2014, Daskalakis2014}. Meanwhile, photon dressing reduces the exciton-exciton interaction strength only by the Hopfield coefficient \cite{Hopfield1958}, which sets the polariton-polariton interaction strength to be \emph{exciton-like}. Along with a much narrower polariton linewidth protected from excitonic inhomogeneous broadening \cite{Houdre1996, Whittaker1998, Diniz2011}, polaritons favor various nonlinear phenomena including the polariton blockade \cite{Verger2006, Munoz-Matutano2019, Delteil2019, Ryou2018, Trivedi2019} without the need to fabricate a nanoscale quantum-dot structure.

Although two principal signatures of BEC, namely, the macroscopic occupation in the ground state and spontaneous emergence of coherence, are routinely observed in polaritonic systems \cite{Deng2002, Kasprzak2006, Balili2007, Christopoulos2007, Daskalakis2014, Plumhof2014}, debates have persisted over whether the observed phenomena can be attributed to BEC or rather to conventional lasing \cite{Butov2012, Deveaud-Pledran2012, Chiocchetta2017}. The former is considered a thermodynamic phase transition, while the latter is governed by dynamic rate equations, although in both cases the underlying formation mechanism is the same---\textcolor{dr}{bosonic} stimulation into a single quantum state \cite{Miesner1998, Ketterle2001}. Only when the polariton gas is in thermal equilibrium can one claim that the relaxation dynamics is not the driving force of \textcolor{dr}{bosonic} stimulation \cite{Snoke2012}. Therefore, full thermalization is one of the most distinguishing features of polariton BEC from its lasing counterpart, and it was long sought in polaritonic systems \cite{Imamoglu1996, Deng2003, Deng2006, Kasprzak2008, Byrnes2014}.

One of the main challenges to reach full thermalization is a short polariton lifetime, which results in a decay of the polariton population before thermal equilibrium is reached. However, when polaritons have a long enough lifetime for thermalization, they show equilibrium properties of BEC such as (1) Bose-Einstein distribution of the state occupation number
\begin{equation}
\mathcal{N}_{\t{BE}}(E) = \frac{1}{e^{(E-\mu)/k_{\t{B}}T}-1}
\label{Eq_NBE}
\end{equation}
with well-defined chemical potential $\mu$ and temperature $T$ (for trapped condensates in quasi-2D) \cite{Sun2017} or (2) power-law decay of spatial and temporal coherence (for freely expanding condensates in 2D) \cite{Caputo2017}.

A Bose-Einstein distribution of polaritons has been considered one of the key criteria for equilibrium BEC \cite{Deng2006, Kasprzak2008, Byrnes2014, Sun2017}, although \textcolor{dr}{quantitatively assessing equilibrium} is experimentally not trivial \cite{Chiocchetta2017}. Even with the experimental demonstration of a Bose-Einstein distribution in a GaAs microcavity \cite{Sun2017}, it is still not fully understood how polaritons thermalize in an optically generated potential. Unlike photon condensates \cite{Klaers2010a, Klaers2010} or organic polariton condensates \cite{Daskalakis2014, Plumhof2014} where fast vibrational relaxation of molecules ($<$\,100~fs) results in sufficient thermalization within their lifetimes, inorganic polariton condensates suffer from slow polariton-phonon scattering ($\sim$10~ps), and they need both self-interactions and interactions with an exciton reservoir for further relaxation. Thus, it is crucial to identify the nature of polariton interactions and scattering rates in various potential landscapes to understand the observed thermalized distribution.

In this paper, we discuss two aspects of the thermalized distribution observed in Ref.~\cite{Sun2017}. We first show that the population distribution is not only determined by relaxation and loss rates, but also by an inhomogeneous pump profile. In particular, a spatially narrow nonresonant pump can inject a pre-thermalized polariton distribution outside the pumped region. We then show how polaritons relax in a trapped region away from the pump, and how the thermalization process can be enhanced through trapping and polariton interactions. The enhancement of the thermalization rate becomes more important in lower-quality systems, e.g., transition metal dichalcogenides in a microcavity, where relaxation toward the ground state is insufficient and the bottleneck effect prevails. One possible route to achieve condensation in such systems would be to facilitate thermalization \cite{Schneider2020} so that polaritons condense before other high-density processes such as exciton-exciton annihilation come into play. In this sense, enhancing thermalization may allow us to overcome the technical limits of material and cavity fabrication and to achieve BEC in novel material systems.
\\

\section{$\mathbf{K}$-space distribution of transported polaritons}\label{Sec:kspace}
Engineering of the potential landscape allows us to control polariton flow and to enhance the effect of polariton interactions \cite{Schneider2016}. One way to create a potential profile for polaritons is to optically pump free carriers and excitons that blueshift the polariton energy through \textcolor{dr}{repulsive} interactions. The optically generated potential can be dynamically controlled, while avoiding etching processes that often lead to unwanted defects. However, complications such as the optical pump rate being coupled to the potential height are added. This results in nontrivial polariton distributions away from a spatially inhomogeneous pump \cite{Cristofolini2018, Ballarini2019}\textcolor{dr}{, where the contributing factors to the distributions are poorly identified.}

\begin{figure*}[bt]
\includegraphics[scale = 1]{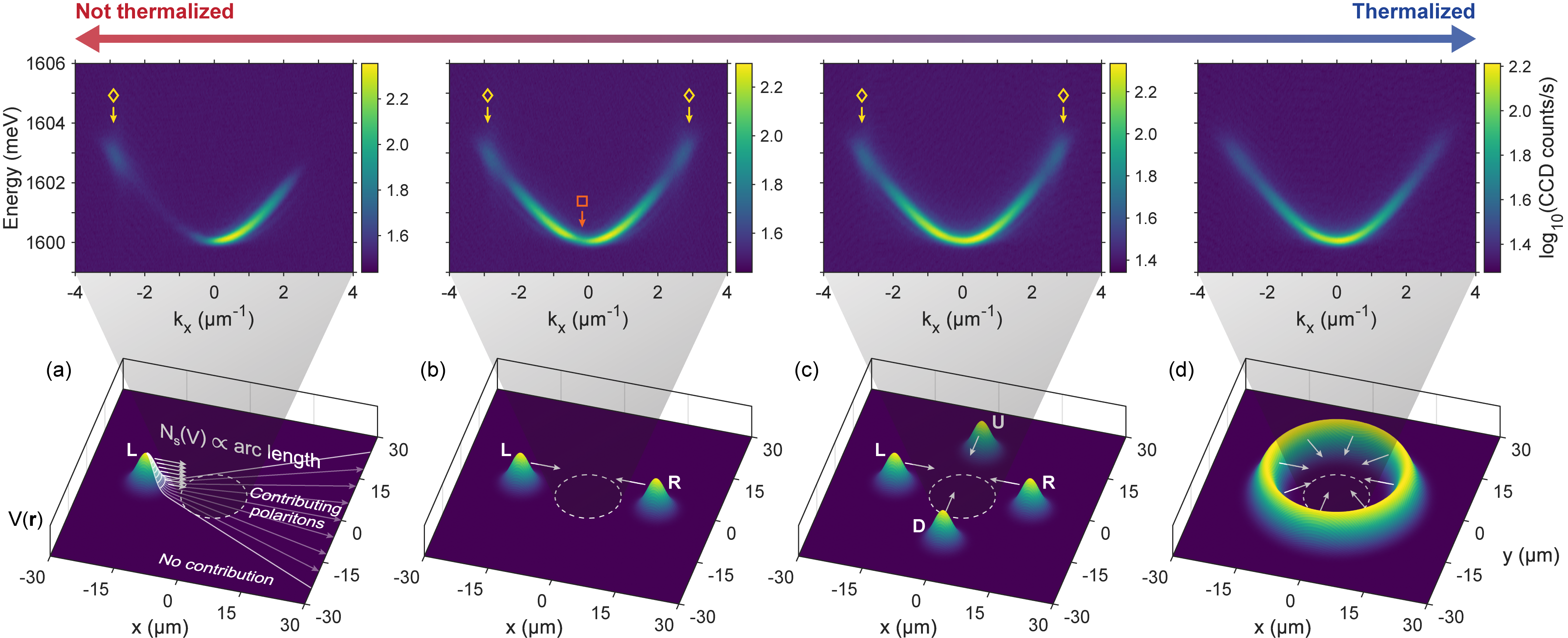}
\caption{Degree of thermalization under various pump geometries at zero detuning and similar pump powers below the condensation threshold. The detection area is shown as a white dashed circle. (a) Polariton distribution under a single Gaussian pump is highly out of equilibrium. The number of source particles $N_{\t{s}}(V)$ from each potential energy contour $V$ that contribute to the detected signal is proportional to the arc length, as indicated by white curves. (b) Two Gaussian pumps yield a distribution similar to but not the same as the sum of (a) and flipped (a). (c) Four Gaussian pumps yield a better-thermalized distribution. (d) Polaritons are fully thermalized in an annular pump geometry. Anomalous spike and dip of population distribution compared to the Bose-Einstein distribution are indicated by yellow diamonds and orange square, respectively. Distributions are plotted on a log scale to emphasize small features.}
\label{Fig1}
\end{figure*}

\begin{figure*}[hbt]
\includegraphics[scale=1]{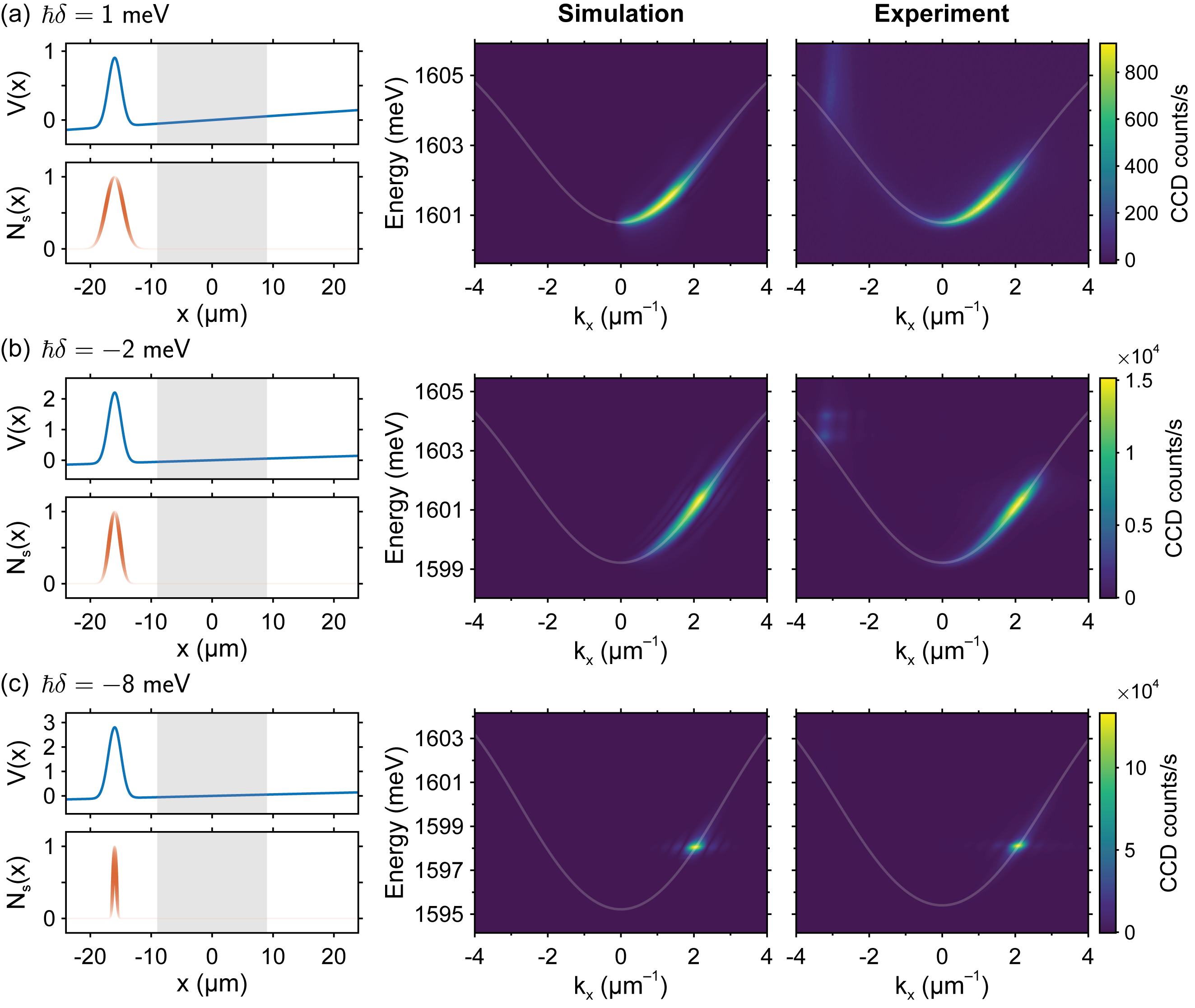}
\caption{\textcolor{dr}{Simulated and measured polariton distributions in the detection area $\abs{\mathbf{r}} < 9~\mu\t{m}$ (gray shaded area) away from a single Gaussian pump beam at $(x, y) = (-16~\mu\t{m}, 0~\mu\t{m})$ for three representative detunings: (a) $\hbar\delta = 1~\t{meV}$, (b) $\hbar\delta = -2~\t{meV}$, and (c) $\hbar\delta = -8~\t{meV}$. For all cases, the pump beam's full width at half maximum (FWHM) was $2.5~\mu\t{m}$ and the pump power was above the condensation threshold. The potential $V(x)$, formed by immobile very-high-$k$ excitons and cavity gradient $6~\mu\t{eV}/\mu\t{m}$, has the same FWHM with that of the pump beam ($\Delta V(x) = 2.5~\mu\t{m}$). Normalized source-particle distributions $N_{\t{s}}(x)$ are plotted as orange lines (the line thickness and color darkness both represent the geometric factor $w(x) \propto x-x_{0}$, where $x_{0}$ is the center position of the pump beam). The FWHM of $N_{\t{s}}(x)$ are (a) $\Delta N_{\text{s}}(x) = 3.3~\mu\text{m}$, (b) $\Delta N_{\text{s}}(x) = 2.1~\mu\text{m}$, and (c) $\Delta N_{\text{s}}(x) = 0.6~\mu\text{m}$. Panels on the right-hand side show simulated and measured polariton distributions.}}
\label{FigS3}
\end{figure*}

Most earlier demonstrations of polariton BEC used a large nonresonant pump spot \cite{Deng2002, Kasprzak2006, Balili2007, Christopoulos2007, Plumhof2014, Daskalakis2014}. In this configuration, polaritons condense where they are created and the condensate barely expands. Even if the condensate expands due to the pump-induced potential hill, it is valid to assume that the condensate particles are predominantly created at the maximum intensity region of the pump spot (with potential energy $V_{\t{max}}$) and then eventually gain momentum $\hbar k_{\t{max}} = \sqrt{2m_{\t{pol}}V_{\t{max}}}$ (where $m_{\t{pol}}$ is the polariton effective mass) corresponding to conversion of potential to kinetic energy \cite{Wertz2010, Christmann2012}.

The polariton distribution below the condensation threshold, however, is more complicated. In particular, its dependence on the pump geometry and cavity detuning is not well understood. Figure~\ref{Fig1} shows four different pump geometries and observed polariton distributions at zero detuning in a GaAs microcavity with polariton lifetime over 200~ps, demonstrated by tracking ballistically propagating polariton wavepackets in space and time \cite{Steger2015, Mukherjee2019}. The single Gaussian pump at $(x, y) = (-16~\mu\t{m}, 0~\mu\t{m})$, shown in Fig.~\ref{Fig1}(a), creates polaritons that are repelled from the pump region. The polariton distribution measured in the detection region (indicated by the white dashed circle, $\abs{\mathbf{r}} < 8~\mu\t{m}$) shows that the average wavevector is $\langle k_{x}\rangle \approx 0.3~\mu\t{m}^{-1}$. The origin of the backscattering peak at $k_{x} = -2.9~\mu\t{m}^{-1}$, indicated by the yellow diamond, is believed to be related to the negative polariton mass at the inflection point \cite{Colas2016, Khamehchi2017}. \textcolor{dr}{The broad energy and momentum distributions are maintained even above the condensation threshold, as shown in Fig.~\ref{FigS3}(a) with similar pump and detection conditions. This behavior is different from the distribution at photonic detuning, shown in Fig.~\ref{FigS3}(c), where most emission is concentrated at very narrow ranges of energy and momentum (centered at $V_{\t{max}}$ and $\hbar k_{\t{max}}$).}

\textcolor{dr}{To describe the detuning-dependent distributions, the multicomponent nature of polaritonic systems must be considered.} We assume that the nonresonant pump creates (1) immobile very-high-$k$ excitons and (2) slightly more mobile inflection-point polaritons \cite{Myers2018}. We call the latter the ``source particles'' because they can directly produce low-$k$ polaritons via parametric scattering, whereas the former particles cannot. The very-high-$k$ excitons form a potential profile $V(\mathbf{r})$ that closely follows the pump intensity profile $I(\mathbf{r})$ due to the small exciton diffusion length. On the other hand, the spatial distribution of the source particles $N_{\t{s}}(\mathbf{r})$ can be broader than that of $I(\mathbf{r})$ through diffusion (this effect is more pronounced with a tightly focused pump beam), while only minimally affecting the overall potential profile. \textcolor{dr}{Thus, there is a mismatch between $V(\mathbf{r})$ and $N_{\t{s}}(\mathbf{r})$, which plays a significant role in the resulting polariton distribution.}

\textcolor{dr}{We simulated the polariton distribution by integrating} contributions from each energy contour $V_{i}$ of the pump-induced potential ($0\leq V_{i} \leq V_{\t{max}}$). In particular, we assumed that polaritons are quasi-thermalized at the generation point owing to the large density of high-$k$ excitons \textcolor{dr}{(the spatially dependent relaxation through polariton-exciton scattering is captured by Wouters \textit{et al}.~\cite{Wouters2010})}, i.e., $\mathcal{N}_{i}(E)$ follows the Bose-Einstein distribution including the pump-induced potential term $V_{i}$. After ballistic transport to the detection area, the resulting distribution $\mathcal{N}_{\t{det}}(E)$ is a weighted sum over all $V_{i}$'s:
\begin{equation}
\mathcal{N}_{\t{det}}(E) = \sum_{i}w_{i}\mathcal{N}_{i}(E) = \sum_{i}\frac{w_{i}}{e^{(E-V_{i} - \mu)/k_{\t{B}}T}-1},
\end{equation}
where the weight $w_{i}$ is proportional to the arc length of each energy contour [white curves in Fig.~\ref{Fig1}(a)].

Figure~\ref{FigS3} shows simulated polariton distributions assuming different diffusion characteristics at different detunings. At an excitonic detuning [Fig.~\ref{FigS3}(a)], the diffusion and broadening of $N_{\t{s}}(\mathbf{r})$ due to repulsive exciton interactions lead to larger weight contributions from $V_{i} < V_{\t{max}}$ than from $V_{i} = V_{\t{max}}$. \textcolor{dr}{The simulated polariton distribution with $\Delta V(\mathbf{r}) = 2.5~\mu\t{m}$ and $\Delta N_{\t{s}}(\mathbf{r}) = 3.3~\mu\t{m}$} shows a broad polariton distribution, in good agreement with the measured distribution. On the other hand, at a photonic detuning [Fig.~\ref{FigS3}(c)], bosonic stimulation starts at the potential maximum and \textcolor{dr}{the majority of polaritons gain momentum $\hbar k_{\t{max}}$ as they propagate to the detection area. The resulting narrow energy and momentum distributions are in line with observations and conventional interpretations with photonic polaritons \cite{Wertz2010, Christmann2012}. Without the mismatch between $V(\mathbf{r})$ and $N_{\t{s}}(\mathbf{r})$, the broad distribution observed at excitonic detunings could not be reproduced at any potential height $V_{\t{max}}$ and chemical potential $\mu$.}

With the two Gaussian pumps shown in Fig.~\ref{Fig1}(b), the distribution seems to be a linear sum of two distributions generated by each pump. However, a careful comparison between the distribution with each individual pump and that with both pumps reveals that nonlinear relaxation processes are present. This indicates that some redistribution of polaritons must follow simple ballistic transport from the pump region to the detection region. Nevertheless, the relaxation is not complete with the two pumps and a population dip at $k_{x} \approx 0~\mu\t{m}^{-1}$ is observed, as indicated by the orange square in Fig.~\ref{Fig1}(b). This is one example of the deviation from the equilibrium Bose-Einstein distribution shown in Eq.~(\ref{Eq_NBE}), where $\mathcal{N}_{\t{BE}}(E)$ monotonically decreases as $E$ increases. The dip disappears when more Gaussian pumps are added, as shown in Fig.~\ref{Fig1}(c). Finally, the backscattering population at $\abs{k_{x}} = 2.9~\mu\t{m}^{-1}$ is pulled down to the ground state when the annular pump shown in Fig.~\ref{Fig1}(d) is used, and the resulting distribution is fully thermalized, i.e., it follows Eq.~(\ref{Eq_NBE}) with well-defined $\mu$ and $T$. These examples show that, even with long-lived polaritons, it is important to choose an appropriate pump geometry and detection configuration to study their equilibrium or nonequilibrium properties \cite{Sun2017, Ballarini2017}.
\\

\section{Real-space distribution of transported polaritons}\label{Sec:rspace}
\begin{figure}[bt]
\includegraphics[scale=1]{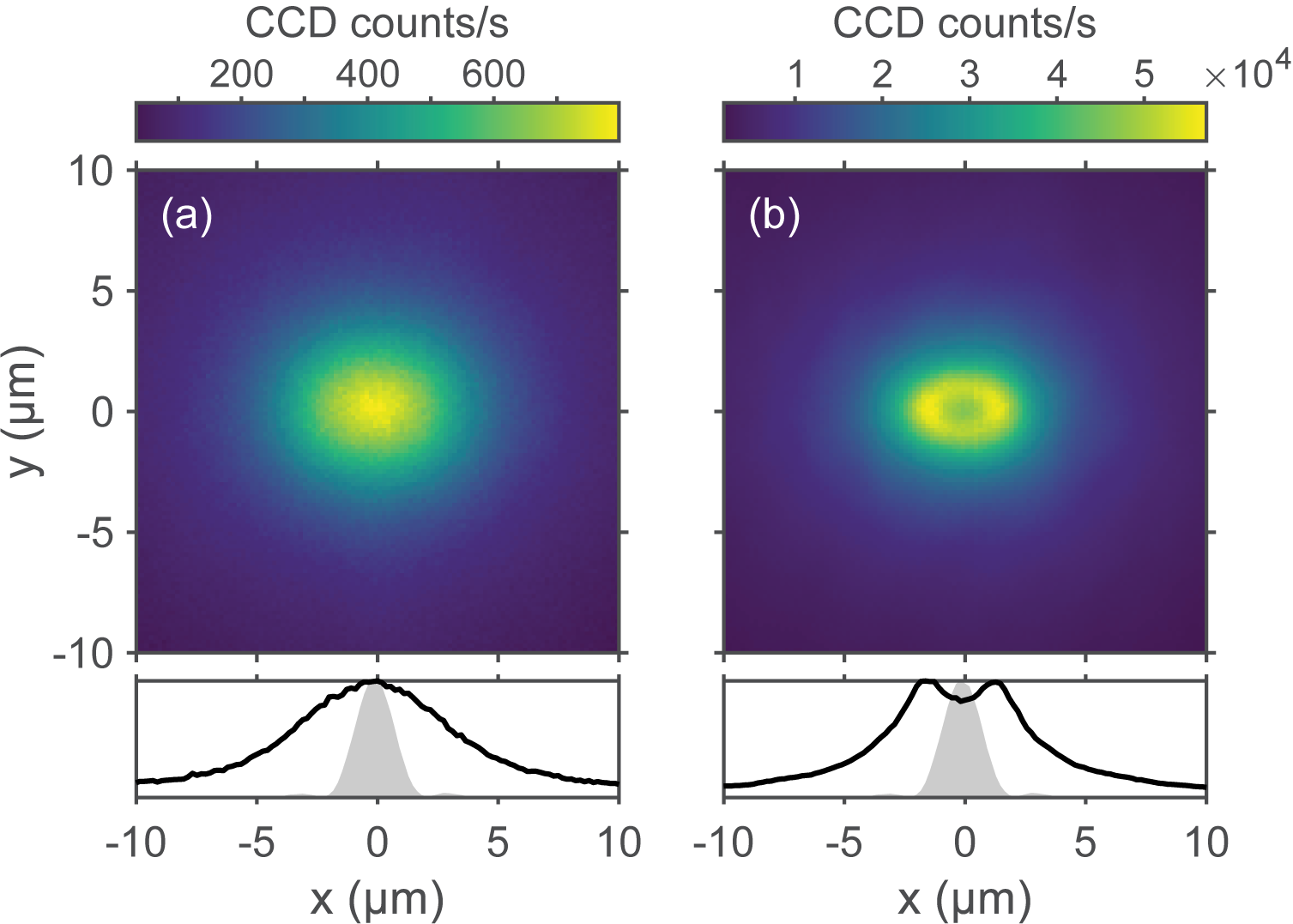}
\caption{Real-space polariton distribution with a single Gaussian pump near zero detuning at (a) a low pump power and (b) a moderate pump power below the condensation threshold. A line cut at $y = 0~\mu\t{m}$ is plotted below as a black line, and the gray shaded area is the laser intensity profile.}
\label{FigS5}
\end{figure}

\textcolor{dr}{The effects of pump geometry and detuning can also be observed in real space, similar to those observed in $k$-space distributions (Figs.~\ref{Fig1} and \ref{FigS3}). With a single Gaussian pump at low power, the real-space emission pattern [Fig.~\ref{FigS5}(a)] is simply governed by the polariton propagation and expansion, $N_{\t{pol}}(\mathbf{r}) \propto (e^{-r/l_{c}})/r$, where $l_{c}$ is the characteristic propagation length.} On the other hand, at moderate pump power just below the condensation threshold, the diffusion of source particles becomes more pronounced by the increased potential gradient, which leads to a halo pattern of polaritons in the real space, as shown in Fig.~\ref{FigS5}(b). \textcolor{dr}{The multicomponent nature and heterogeneity in diffusion lengths can explain the halo patterns observed at high excitation densities. This explanation may extend to other excitonic systems where similar halo patterns are observed \cite{Rivera2016,Kulig2018,PereaCausin2019}; repulsive interactions cannot solely explain the pattern if there is only one type of particles. }

\begin{figure}[bt]
\includegraphics[scale=1]{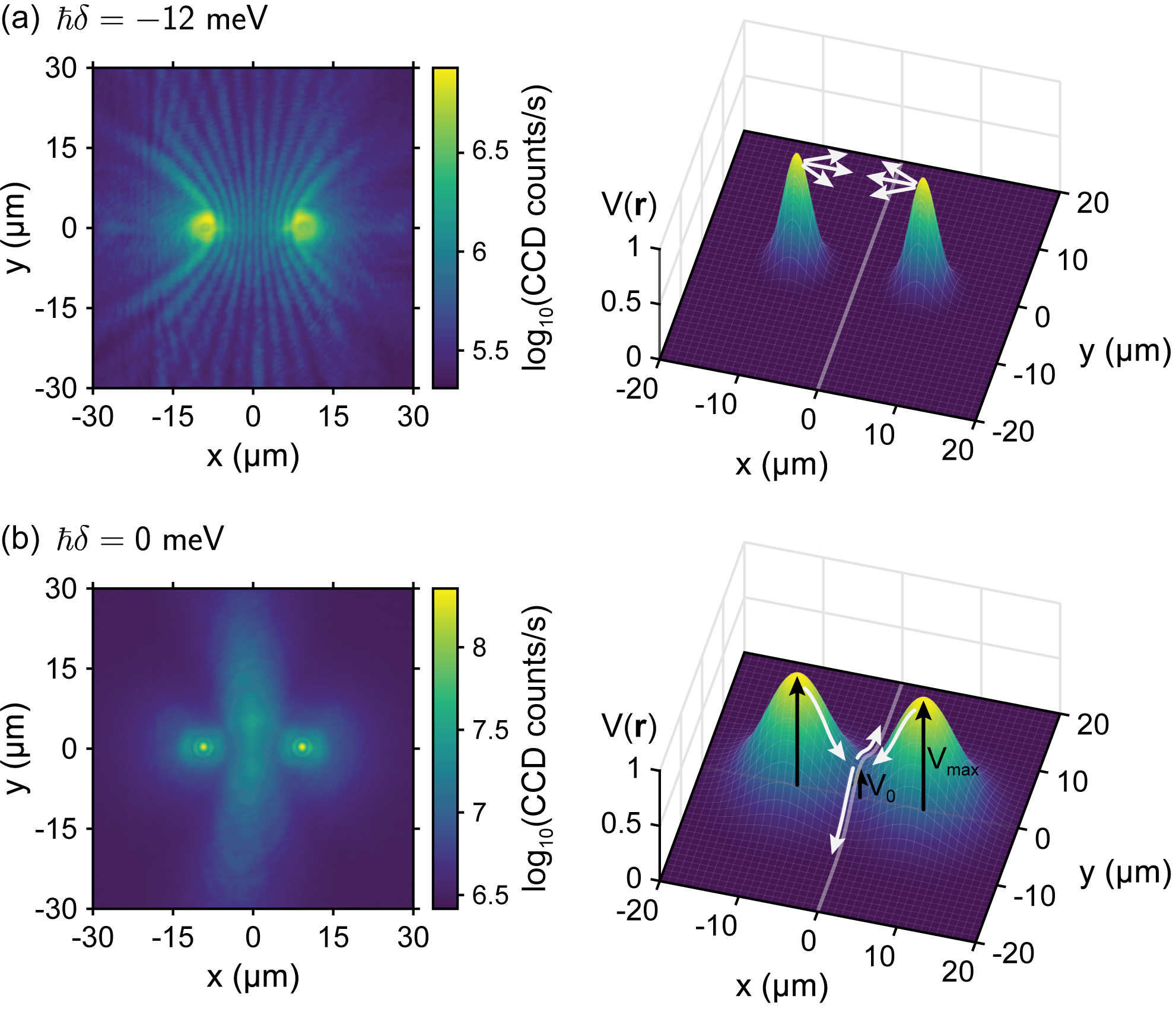}
\caption{Real-space polariton distribution with two Gaussian pumps above the condensation threshold at (a) $\hbar\delta = -12~\t{meV}$ and (b) $\hbar\delta = 0~\t{meV}$. The potential profile $V(\mathbf{r})$ in each case is illustrated schematically on the right, where the polariton propagation directions are shown with white arrows. The distance between the two pumps is $18.5~\mu\t{m}$.}
\label{FigS6}
\end{figure}

\textcolor{dr}{With two Gaussian pump beams located at $(x, y) = (\pm 9.2~\mu\t{m}, 0~\mu\t{m})$, the effects of detuning show up more dramatically in real space.} At a photonic detuning [Fig.~\ref{FigS6}(a), $\hbar\delta = -12~\t{meV}$], most polaritons are generated near the top of the potential hill $V_{\t{max}}$ and expand away from the two generation points, resulting in an interference pattern with $k_{x} = \pm\sqrt{2m_{\t{pol}}V_{\t{max}}}/\hbar$ components, similar to that reported by Tosi \textit{et al}.\ \cite{Tosi2012a}. In stark contrast, at a zero detuning [Fig.~\ref{FigS6}(b), $\hbar\delta = 0~\t{meV}$], the $k_{x} = \pm\sqrt{2m_{\t{pol}}V_{\t{max}}}/\hbar$ components disappear and polaritons dominantly flow along the line $x = 0$, perpendicular to the line  connecting two pumps ($y = 0$). After analyzing the potential landscape using energy-resolved real-space photoluminescence measurements with a $k$-space filter (that selects only $k < 0.5~\mu\t{m}^{-1}$), we conclude that this peculiar pattern arises from the broadened potential profile $V(\mathbf{r})$ and even broader source-particle distribution $N_{\t{s}}(\mathbf{r})$. The broadened potential profiles of the two pumped regions merge at the center $(x, y) = (0~\mu\t{m}, 0~\mu\t{m})$, raising the local potential energy to $V_{0}$ \textcolor{dr}{to form a saddle point}. If the majority of polaritons are created from the top of the potential hills, then they are still expected to retain its wavevector components $k_{x} = \pm \sqrt{2m_{\t{pol}}(V_{\t{max}} - V_{0})}/\hbar$ \textcolor{dr}{and interfere} at the center. However, an even broader $N_{\t{s}}(\mathbf{r})$ profile leads to the generation of polaritons directly from the center, which then gain only $k_{y} = \pm \sqrt{2m_{\t{pol}}V_{0}}/\hbar$ components. This interpretation gives a simple explanation to the seemingly puzzling polariton behavior observed between nonresonant pumps \cite{Dreismann2014, Cristofolini2018}.
\\

\section{Redistribution by trapping and interactions}
To elucidate the mechanism of full thermalization in the annular trap, we instead used four Gaussian pumps in a geometry similar to that in Fig.~\ref{Fig1}(c) and analyzed the contributions from each pump. This allows us to deconvolute which parts of the trapped distribution can arise from a linear sum, and to monitor how much population is nonlinearly redistributed in the two-dimensional $k$ space. The four pumps are labeled as left (L), right (R), up (U), and down (D), indicating their positions relative to the origin $(x, y) = (0, 0)$. The potential formed by the four pumps can effectively trap polaritons (even though there are four diagonal leakage channels), while each pump alone cannot trap polaritons. Therefore, by comparing all four pumps (trapped) and individual pumps (untrapped), the effect of trapping can be investigated.

\begin{figure*}[bt]
\includegraphics[scale = 1]{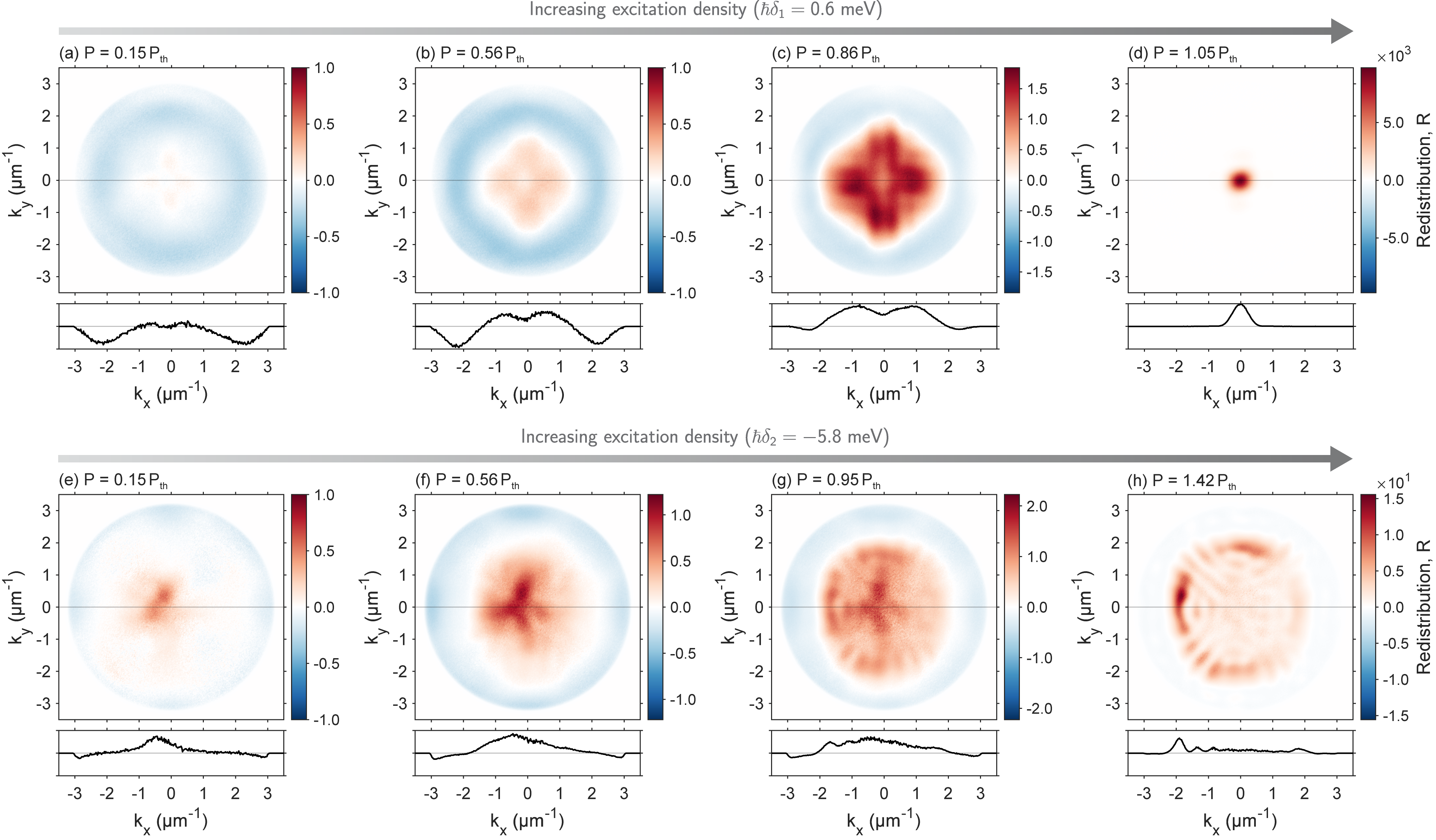}
\caption{Redistribution function $R(\mathbf{k})$, defined in Eq.~(\ref{R}), at four representative pump powers at detuning (a-d) $\hbar\delta_{1} = 0.6~\t{meV}$ and (e-h) $\hbar\delta_{2} = -5.8~\t{meV}$.}
\label{Fig2}
\end{figure*}

We measured polariton distribution in the two-dimensional $k$ space with all four pumps, which we denote as $N_{\text{LRUD}}(\mathbf{k})$, and compared it with the sum of distributions measured individually with each pump, i.e., $\sum_{i}N_{i}(\mathbf{k})$, where $i = \text{L, R, U, and D}$. We define a normalized redistribution function $R(\mathbf{k})$ that quantifies the extent of the nonlinear gain or loss:
\begin{equation}
R \equiv \frac{N_{\text{LRUD}} - \left(N_{\text{L}} + N_{\text{R}} + N_{\text{U}} + N_{\text{D}}\right)}{N_{\text{L}} + N_{\text{R}} + N_{\text{U}} + N_{\text{D}}}.
\label{R}
\end{equation}
If there are no interactions between polaritons and if the trapping potential has no effect on the relaxation, then $N_{\t{LRUD}} = N_{\t{L}} + N_{\t{R}} + N_{\t{U}} + N_{\t{D}}$ and the redistribution function must be zero. If polaritons interact and exchange momentum as described by the term $\mathcal{H}_{\t{int}} = \hat{a}_{\mathbf{k}_{1}+\mathbf{q}}^{\dagger}\hat{a}_{\mathbf{k}_{2}-\mathbf{q}}^{\dagger}\hat{a}_{\mathbf{k}_{1}}\hat{a}_{\mathbf{k}_{2}}$, where $\hat{a}_{\mathbf{k}}$ is the annihilation operator for a polariton at $\mathbf{k}$, then $R$ deviates from zero. \textcolor{dr}{For example, $R > 0$ means that a larger number of polaritons are detected if they are trapped (compared to the untrapped case), i.e., there is nonlinear gain by trapping. Similarly, $R < 0$ means that there are less polaritons if they are trapped (compared to the untrapped case), i.e., there is nonlinear loss by trapping.}


A series of redistribution functions $R(\mathbf{k})$ with different excitation densities at near-zero detuning ($\hbar\delta_{1} = 0.6~\t{meV}$) is shown in Figs.~\ref{Fig2}(a-d). Surprisingly, even at a low pump power ($P = 0.15 P_{\t{th}}$) well below the condensation threshold power $P_{\t{th}}$, a sizable amount of polariton depletion ($R < 0$) is observed at $\abs{\mathbf{k}} = 2.3~\mu\t{m}^{-1}$, where the inflection point of the polariton dispersion is located. At this so-called ``magic'' angle, the parametric polariton-polariton scattering is facilitated due to energy-momentum matching \cite{Savvidis2000a}. This shows that, in the presence of the trapping potential, polaritons can more efficiently scatter out of the inflection point to other $k$ states, some of which are outside the numerical aperture (NA = 0.42) of the objective lens.

At intermediate powers ($0.3P_{\t{th}} < P < 0.8P_{\t{th}}$), the low-energy states ($\abs{\mathbf{k}} < 1~\mu\t{m}^{-1}$) show $R > 0$, while the inflection-point polaritons show $R < 0$. This indicates that the parametric polariton-polariton scattering process \cite{Savvidis2000a} becomes more efficient in the presence of trapping potential. It turns out that the most probable final state for this process is not the polariton ground state but rather the $\abs{\mathbf{k}} \approx 0.5~\mu\t{m}^{-1}$ state \cite{Porras2002}, which is determined by both the shape of the dispersion and the population distribution of high-$k$ excitons (see Fig.~S2 for details).

Above the condensation threshold ($P > P_{\t{th}}$), bosonic stimulation into the trap's ground state at $\mathbf{k} = 0$ is observed, while the stimulation does not start in the untrapped case at the same polariton density. By forming a trapping potential, the density of states $\mathcal{D}(E)$ is reduced so that the occupation number $\mathcal{N}(E)$ \emph{per state} is larger, even though the total polariton density $n = (1/A)\int \mathcal{D}(E)\mathcal{N}(E)dE$ remains the same, where $A$ is the system area. The bosonic stimulation condition $\mathcal{N}(E)>1$ is more easily reached by the trapped ground state, \textcolor{dr}{which is selected as the condensate state}.

\textcolor{dr}{The trapping facilitates the redistribution of polaritons from high- to low-energy states by providing polaritons more time to interact}, as directly visualized by Rozas \textit{et al}.\ \cite{Rozas2019}. \textcolor{dr}{Freely} expanding polaritons, in contrast, have to rely \textcolor{dr}{only} on polariton-phonon interactions to relax down to the ground state \cite{Ballarini2017}. \textcolor{dr}{Our} finding also suggests that high-$k$ excitons (outside the objective lens NA) are more efficiently generated in an annular trap \textcolor{dr}{than without a trap. This is more evident in the low-density data shown in Fig.~\ref{Fig2}(a), where the population in the depletion region ($\abs{\mathbf{k}} \approx 2.3~\mu\t{m}^{-1}$) is not fully redistributed to the lower-$k$ region but mostly to the higher-$k$ region that are not detected. This effect may have affected the measurement of the polariton-polariton interaction strength in Ref.~\cite{Sun2017a} by promoting the high-$k$ dark exciton generation in a large annular trap.}

\begin{figure}[bt]
\includegraphics[scale = 1]{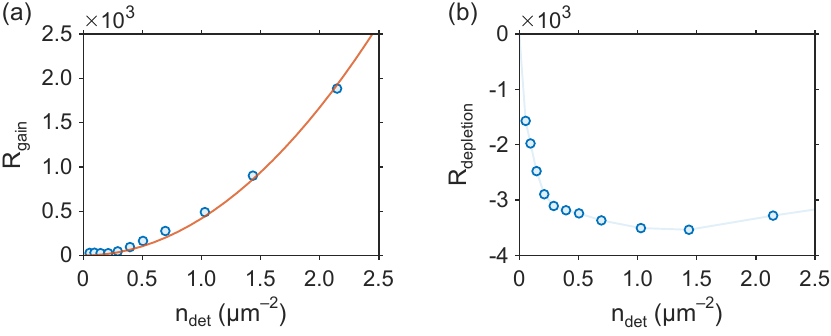}
\caption{\textcolor{dr}{Integrated redistribution functions at detuning $\hbar\delta_{1} = 0.6~\t{meV}$ as a function of the polariton density in the detection area ($n_{\t{det}}$) up to just below the condensation threshold. (a) $R_{\t{gain}}$ is the redistribution function integrated within the gain region ($\abs{\mathbf{k}} < 0.5~\mu\t{m}^{-1}$). The orange line is a quadratic fit. (b) $R_{\t{depletion}}$ is the redistribution function integrated within the depletion region ($2.1~\mu\t{m}^{-1} < \abs{\mathbf{k}} < 2.5~\mu\t{m}^{-1}$).}}
\label{Fig3}
\end{figure}

To investigate the role of polariton interactions, we plotted density-dependent $R(\mathbf{k})$ integrated within the gain region ($\abs{\mathbf{k}}<0.5~\mu\t{m}^{-1}$) in Fig.~\ref{Fig3}(a). The quadratic scaling of the effective redistribution rate with the polariton density suggests that polariton-polariton or polariton-exciton scattering processes dominate over other processes including polariton-phonon scattering---relaxation via phonon scattering is expected to show a linear scaling with the polariton density. The integrated $R(\mathbf{k})$ within the depletion region ($2.1~\mu\t{m}^{-1}<\abs{\mathbf{k}}<2.5~\mu\t{m}^{-1}$), plotted in Fig.~\ref{Fig3}(b), shows that the depletion rate saturates at higher densities. The depletion at the inflection point of polariton dispersion strongly supports the important role of parametric polariton-polariton scattering during relaxation below the condensation threshold.

We varied the detuning to $\hbar\delta_{2} = -5.8~\t{meV}$ so that the polariton interaction strength is weaker [Figs.~\ref{Fig2}(e-h)]. At this photonic detuning, the gain region ($R>0$) is shifted from the origin ($\mathbf{k} = 0$) due to the cavity gradient and increased polariton mobility. The depletion region ($R<0$) is observed at $\abs{\mathbf{k}} = 3.0~\mu\t{m}^{-1}$, which corresponds to the inflection point at this detuning (see Fig.~S8). Above the condensation threshold, bosonic stimulation occurs at propagating modes with a finite wavevector $k_{\t{p}}$, determined by the pump-induced potential height $V_{\t{max}}$, in stark contrast to the zero-detuning case [Figs.~\ref{Fig2}(a-d)]. Weakly interacting polaritons cannot easily relax down to the ground state; they form a nonequilibrium condensate in the excited state.
\\

\section{Conclusions}
We demonstrated enhancement of the thermalization rate in the presence of a trapping potential and polariton interactions. The polariton redistribution is quantified by comparing the polariton distribution under four nonresonant Gaussian pumps in a square geometry and the distributions measured individually with each pump. Significant depletion of inflection-point polaritons as well as density- and detuning-dependent redistribution rates reveal that parametric polariton-polariton scattering is responsible for the enhanced thermalization in optically generated potentials. Our work paves the way to overcome the bottleneck effect and to achieve polariton BEC in low-quality microcavities. Furthermore, understanding polariton distributions and their interactions away from the pump allows us to predict which trapped state is dominantly injected\textcolor{dr}{, which determines the coupling} between the pump sites in polariton graph simulators \cite{Cristofolini2013, Berloff2017, Sun2018}.
\\

\begin{acknowledgments}
Y.Y., J.D., and K.A.N.\ were supported in part by Skoltech as part of the Skoltech-MIT Next Generation Program. J.D.\ acknowledges partial support from the NSERC-PGS program. M.S.\ and D.W.S.\ were supported by the National Science Foundation under Grant Number DMR-2004570. K.W.W.\ and L.N.P.\ were partially funded by the Gordon and Betty Moore Foundation through the EPiQS initiative Grant GBMF4420 and by the National Science Foundation MRSEC Grant DMR-1420541.
\\
\end{acknowledgments}

\appendix
\renewcommand{\thesubsection}{\thesection.\arabic{subsection}}
\makeatletter
\renewcommand{\p@subsection}{}
\renewcommand{\p@subsubsection}{}
\makeatother


\begin{thebibliography}{58}%
\makeatletter
\providecommand \@ifxundefined [1]{%
 \@ifx{#1\undefined}
}%
\providecommand \@ifnum [1]{%
 \ifnum #1\expandafter \@firstoftwo
 \else \expandafter \@secondoftwo
 \fi
}%
\providecommand \@ifx [1]{%
 \ifx #1\expandafter \@firstoftwo
 \else \expandafter \@secondoftwo
 \fi
}%
\providecommand \natexlab [1]{#1}%
\providecommand \enquote  [1]{``#1''}%
\providecommand \bibnamefont  [1]{#1}%
\providecommand \bibfnamefont [1]{#1}%
\providecommand \citenamefont [1]{#1}%
\providecommand \href@noop [0]{\@secondoftwo}%
\providecommand \href [0]{\begingroup \@sanitize@url \@href}%
\providecommand \@href[1]{\@@startlink{#1}\@@href}%
\providecommand \@@href[1]{\endgroup#1\@@endlink}%
\providecommand \@sanitize@url [0]{\catcode `\\12\catcode `\$12\catcode
  `\&12\catcode `\#12\catcode `\^12\catcode `\_12\catcode `\%12\relax}%
\providecommand \@@startlink[1]{}%
\providecommand \@@endlink[0]{}%
\providecommand \url  [0]{\begingroup\@sanitize@url \@url }%
\providecommand \@url [1]{\endgroup\@href {#1}{\urlprefix }}%
\providecommand \urlprefix  [0]{URL }%
\providecommand \Eprint [0]{\href }%
\providecommand \doibase [0]{https://doi.org/}%
\providecommand \selectlanguage [0]{\@gobble}%
\providecommand \bibinfo  [0]{\@secondoftwo}%
\providecommand \bibfield  [0]{\@secondoftwo}%
\providecommand \translation [1]{[#1]}%
\providecommand \BibitemOpen [0]{}%
\providecommand \bibitemStop [0]{}%
\providecommand \bibitemNoStop [0]{.\EOS\space}%
\providecommand \EOS [0]{\spacefactor3000\relax}%
\providecommand \BibitemShut  [1]{\csname bibitem#1\endcsname}%
\let\auto@bib@innerbib\@empty
\bibitem [{\citenamefont {Yamamoto}\ and\ \citenamefont
  {\.{I}mamo\u{g}lu}(1999)}]{Yamamoto1999}%
  \BibitemOpen
  \bibfield  {author} {\bibinfo {author} {\bibfnamefont {Y.}~\bibnamefont
  {Yamamoto}}\ and\ \bibinfo {author} {\bibfnamefont {A.}~\bibnamefont
  {\.{I}mamo\u{g}lu}},\ }\href
  {https://www.wiley.com/en-us/Mesoscopic+Quantum+Optics-p-9780471148746}
  {\emph {\bibinfo {title} {Mesoscopic Quantum Optics}}}\ (\bibinfo
  {publisher} {John Wiley \& Sons},\ \bibinfo {year} {1999})\BibitemShut
  {NoStop}%
\bibitem [{\citenamefont {Deng}\ \emph {et~al.}(2010)\citenamefont {Deng},
  \citenamefont {Haug},\ and\ \citenamefont {Yamamoto}}]{Deng2010}%
  \BibitemOpen
  \bibfield  {author} {\bibinfo {author} {\bibfnamefont {H.}~\bibnamefont
  {Deng}}, \bibinfo {author} {\bibfnamefont {H.}~\bibnamefont {Haug}},\ and\
  \bibinfo {author} {\bibfnamefont {Y.}~\bibnamefont {Yamamoto}},\ }\href
  {https://doi.org/10.1103/RevModPhys.82.1489} {\bibfield  {journal} {\bibinfo
  {journal} {Rev. Mod. Phys.}\ }\textbf {\bibinfo {volume} {82}},\ \bibinfo
  {pages} {1489} (\bibinfo {year} {2010})}\BibitemShut {NoStop}%
\bibitem [{\citenamefont {Carusotto}\ and\ \citenamefont
  {Ciuti}(2013)}]{Carusotto2013}%
  \BibitemOpen
  \bibfield  {author} {\bibinfo {author} {\bibfnamefont {I.}~\bibnamefont
  {Carusotto}}\ and\ \bibinfo {author} {\bibfnamefont {C.}~\bibnamefont
  {Ciuti}},\ }\href {https://doi.org/10.1103/RevModPhys.85.299} {\bibfield
  {journal} {\bibinfo  {journal} {Rev. Mod. Phys.}\ }\textbf {\bibinfo {volume}
  {85}},\ \bibinfo {pages} {299} (\bibinfo {year} {2013})}\BibitemShut
  {NoStop}%
\bibitem [{\citenamefont {Deng}\ \emph {et~al.}(2002)\citenamefont {Deng},
  \citenamefont {Weihs}, \citenamefont {Santori}, \citenamefont {Bloch},\ and\
  \citenamefont {Yamamoto}}]{Deng2002}%
  \BibitemOpen
  \bibfield  {author} {\bibinfo {author} {\bibfnamefont {H.}~\bibnamefont
  {Deng}}, \bibinfo {author} {\bibfnamefont {G.}~\bibnamefont {Weihs}},
  \bibinfo {author} {\bibfnamefont {C.}~\bibnamefont {Santori}}, \bibinfo
  {author} {\bibfnamefont {J.}~\bibnamefont {Bloch}},\ and\ \bibinfo {author}
  {\bibfnamefont {Y.}~\bibnamefont {Yamamoto}},\ }\href
  {https://doi.org/10.1126/science.1074464} {\bibfield  {journal} {\bibinfo
  {journal} {Science}\ }\textbf {\bibinfo {volume} {298}},\ \bibinfo {pages}
  {199} (\bibinfo {year} {2002})}\BibitemShut {NoStop}%
\bibitem [{\citenamefont {Kasprzak}\ \emph {et~al.}(2006)\citenamefont
  {Kasprzak}, \citenamefont {Richard}, \citenamefont {Kundermann},
  \citenamefont {Baas}, \citenamefont {Jeambrun}, \citenamefont {Keeling},
  \citenamefont {Marchetti}, \citenamefont {Szymanska}, \citenamefont
  {Andr{\'e}}, \citenamefont {Staehli}, \citenamefont {Savona}, \citenamefont
  {Littlewood}, \citenamefont {Deveaud},\ and\ \citenamefont
  {Dang}}]{Kasprzak2006}%
  \BibitemOpen
  \bibfield  {author} {\bibinfo {author} {\bibfnamefont {J.}~\bibnamefont
  {Kasprzak}}, \bibinfo {author} {\bibfnamefont {M.}~\bibnamefont {Richard}},
  \bibinfo {author} {\bibfnamefont {S.}~\bibnamefont {Kundermann}}, \bibinfo
  {author} {\bibfnamefont {A.}~\bibnamefont {Baas}}, \bibinfo {author}
  {\bibfnamefont {P.}~\bibnamefont {Jeambrun}}, \bibinfo {author}
  {\bibfnamefont {J.~M.~J.}\ \bibnamefont {Keeling}}, \bibinfo {author}
  {\bibfnamefont {F.~M.}\ \bibnamefont {Marchetti}}, \bibinfo {author}
  {\bibfnamefont {M.~H.}\ \bibnamefont {Szymanska}}, \bibinfo {author}
  {\bibfnamefont {R.}~\bibnamefont {Andr{\'e}}}, \bibinfo {author}
  {\bibfnamefont {J.~L.}\ \bibnamefont {Staehli}}, \bibinfo {author}
  {\bibfnamefont {V.}~\bibnamefont {Savona}}, \bibinfo {author} {\bibfnamefont
  {P.~B.}\ \bibnamefont {Littlewood}}, \bibinfo {author} {\bibfnamefont
  {B.}~\bibnamefont {Deveaud}},\ and\ \bibinfo {author} {\bibfnamefont {L.~S.}\
  \bibnamefont {Dang}},\ }\href {https://doi.org/10.1038/nature05131}
  {\bibfield  {journal} {\bibinfo  {journal} {Nature}\ }\textbf {\bibinfo
  {volume} {443}},\ \bibinfo {pages} {409} (\bibinfo {year}
  {2006})}\BibitemShut {NoStop}%
\bibitem [{\citenamefont {Balili}\ \emph {et~al.}(2007)\citenamefont {Balili},
  \citenamefont {Hartwell}, \citenamefont {Snoke}, \citenamefont {Pfeiffer},\
  and\ \citenamefont {West}}]{Balili2007}%
  \BibitemOpen
  \bibfield  {author} {\bibinfo {author} {\bibfnamefont {R.}~\bibnamefont
  {Balili}}, \bibinfo {author} {\bibfnamefont {V.}~\bibnamefont {Hartwell}},
  \bibinfo {author} {\bibfnamefont {D.}~\bibnamefont {Snoke}}, \bibinfo
  {author} {\bibfnamefont {L.}~\bibnamefont {Pfeiffer}},\ and\ \bibinfo
  {author} {\bibfnamefont {K.}~\bibnamefont {West}},\ }\href
  {https://doi.org/10.1126/science.1140990} {\bibfield  {journal} {\bibinfo
  {journal} {Science}\ }\textbf {\bibinfo {volume} {316}},\ \bibinfo {pages}
  {1007} (\bibinfo {year} {2007})}\BibitemShut {NoStop}%
\bibitem [{\citenamefont {Christopoulos}\ \emph {et~al.}(2007)\citenamefont
  {Christopoulos}, \citenamefont {von H\"ogersthal}, \citenamefont {Grundy},
  \citenamefont {Lagoudakis}, \citenamefont {Kavokin}, \citenamefont
  {Baumberg}, \citenamefont {Christmann}, \citenamefont {Butt\'e},
  \citenamefont {Feltin}, \citenamefont {Carlin},\ and\ \citenamefont
  {Grandjean}}]{Christopoulos2007}%
  \BibitemOpen
  \bibfield  {author} {\bibinfo {author} {\bibfnamefont {S.}~\bibnamefont
  {Christopoulos}}, \bibinfo {author} {\bibfnamefont {G.~B.~H.}\ \bibnamefont
  {von H\"ogersthal}}, \bibinfo {author} {\bibfnamefont {A.~J.~D.}\
  \bibnamefont {Grundy}}, \bibinfo {author} {\bibfnamefont {P.~G.}\
  \bibnamefont {Lagoudakis}}, \bibinfo {author} {\bibfnamefont {A.~V.}\
  \bibnamefont {Kavokin}}, \bibinfo {author} {\bibfnamefont {J.~J.}\
  \bibnamefont {Baumberg}}, \bibinfo {author} {\bibfnamefont {G.}~\bibnamefont
  {Christmann}}, \bibinfo {author} {\bibfnamefont {R.}~\bibnamefont {Butt\'e}},
  \bibinfo {author} {\bibfnamefont {E.}~\bibnamefont {Feltin}}, \bibinfo
  {author} {\bibfnamefont {J.-F.}\ \bibnamefont {Carlin}},\ and\ \bibinfo
  {author} {\bibfnamefont {N.}~\bibnamefont {Grandjean}},\ }\href
  {https://doi.org/10.1103/PhysRevLett.98.126405} {\bibfield  {journal}
  {\bibinfo  {journal} {Phys. Rev. Lett.}\ }\textbf {\bibinfo {volume} {98}},\
  \bibinfo {pages} {126405} (\bibinfo {year} {2007})}\BibitemShut {NoStop}%
\bibitem [{\citenamefont {Plumhof}\ \emph {et~al.}(2014)\citenamefont
  {Plumhof}, \citenamefont {Stöferle}, \citenamefont {Mai}, \citenamefont
  {Scherf},\ and\ \citenamefont {Mahrt}}]{Plumhof2014}%
  \BibitemOpen
  \bibfield  {author} {\bibinfo {author} {\bibfnamefont {J.~D.}\ \bibnamefont
  {Plumhof}}, \bibinfo {author} {\bibfnamefont {T.}~\bibnamefont {Stöferle}},
  \bibinfo {author} {\bibfnamefont {L.}~\bibnamefont {Mai}}, \bibinfo {author}
  {\bibfnamefont {U.}~\bibnamefont {Scherf}},\ and\ \bibinfo {author}
  {\bibfnamefont {R.~F.}\ \bibnamefont {Mahrt}},\ }\href
  {https://doi.org/10.1038/nmat3825} {\bibfield  {journal} {\bibinfo  {journal}
  {Nat. Mater.}\ }\textbf {\bibinfo {volume} {13}},\ \bibinfo {pages} {247}
  (\bibinfo {year} {2014})}\BibitemShut {NoStop}%
\bibitem [{\citenamefont {Daskalakis}\ \emph {et~al.}(2014)\citenamefont
  {Daskalakis}, \citenamefont {Maier}, \citenamefont {Murray},\ and\
  \citenamefont {K{\'{e}}na-Cohen}}]{Daskalakis2014}%
  \BibitemOpen
  \bibfield  {author} {\bibinfo {author} {\bibfnamefont {K.~S.}\ \bibnamefont
  {Daskalakis}}, \bibinfo {author} {\bibfnamefont {S.~A.}\ \bibnamefont
  {Maier}}, \bibinfo {author} {\bibfnamefont {R.}~\bibnamefont {Murray}},\ and\
  \bibinfo {author} {\bibfnamefont {S.}~\bibnamefont {K{\'{e}}na-Cohen}},\
  }\href {https://doi.org/10.1038/nmat3874} {\bibfield  {journal} {\bibinfo
  {journal} {Nat. Mater.}\ }\textbf {\bibinfo {volume} {13}},\ \bibinfo {pages}
  {271} (\bibinfo {year} {2014})}\BibitemShut {NoStop}%
\bibitem [{\citenamefont {Hopfield}(1958)}]{Hopfield1958}%
  \BibitemOpen
  \bibfield  {author} {\bibinfo {author} {\bibfnamefont {J.~J.}\ \bibnamefont
  {Hopfield}},\ }\href {https://doi.org/10.1103/PhysRev.112.1555} {\bibfield
  {journal} {\bibinfo  {journal} {Phys. Rev.}\ }\textbf {\bibinfo {volume}
  {112}},\ \bibinfo {pages} {1555} (\bibinfo {year} {1958})}\BibitemShut
  {NoStop}%
\bibitem [{\citenamefont {Houdr\'e}\ \emph {et~al.}(1996)\citenamefont
  {Houdr\'e}, \citenamefont {Stanley},\ and\ \citenamefont
  {Ilegems}}]{Houdre1996}%
  \BibitemOpen
  \bibfield  {author} {\bibinfo {author} {\bibfnamefont {R.}~\bibnamefont
  {Houdr\'e}}, \bibinfo {author} {\bibfnamefont {R.~P.}\ \bibnamefont
  {Stanley}},\ and\ \bibinfo {author} {\bibfnamefont {M.}~\bibnamefont
  {Ilegems}},\ }\href {https://doi.org/10.1103/PhysRevA.53.2711} {\bibfield
  {journal} {\bibinfo  {journal} {Phys. Rev. A}\ }\textbf {\bibinfo {volume}
  {53}},\ \bibinfo {pages} {2711} (\bibinfo {year} {1996})}\BibitemShut
  {NoStop}%
\bibitem [{\citenamefont {Whittaker}(1998)}]{Whittaker1998}%
  \BibitemOpen
  \bibfield  {author} {\bibinfo {author} {\bibfnamefont {D.~M.}\ \bibnamefont
  {Whittaker}},\ }\href {https://doi.org/10.1103/PhysRevLett.80.4791}
  {\bibfield  {journal} {\bibinfo  {journal} {Phys. Rev. Lett.}\ }\textbf
  {\bibinfo {volume} {80}},\ \bibinfo {pages} {4791} (\bibinfo {year}
  {1998})}\BibitemShut {NoStop}%
\bibitem [{\citenamefont {Diniz}\ \emph {et~al.}(2011)\citenamefont {Diniz},
  \citenamefont {Portolan}, \citenamefont {Ferreira}, \citenamefont {G\'erard},
  \citenamefont {Bertet},\ and\ \citenamefont {Auff\`eves}}]{Diniz2011}%
  \BibitemOpen
  \bibfield  {author} {\bibinfo {author} {\bibfnamefont {I.}~\bibnamefont
  {Diniz}}, \bibinfo {author} {\bibfnamefont {S.}~\bibnamefont {Portolan}},
  \bibinfo {author} {\bibfnamefont {R.}~\bibnamefont {Ferreira}}, \bibinfo
  {author} {\bibfnamefont {J.~M.}\ \bibnamefont {G\'erard}}, \bibinfo {author}
  {\bibfnamefont {P.}~\bibnamefont {Bertet}},\ and\ \bibinfo {author}
  {\bibfnamefont {A.}~\bibnamefont {Auff\`eves}},\ }\href
  {https://doi.org/10.1103/PhysRevA.84.063810} {\bibfield  {journal} {\bibinfo
  {journal} {Phys. Rev. A}\ }\textbf {\bibinfo {volume} {84}},\ \bibinfo
  {pages} {063810} (\bibinfo {year} {2011})}\BibitemShut {NoStop}%
\bibitem [{\citenamefont {Verger}\ \emph {et~al.}(2006)\citenamefont {Verger},
  \citenamefont {Ciuti},\ and\ \citenamefont {Carusotto}}]{Verger2006}%
  \BibitemOpen
  \bibfield  {author} {\bibinfo {author} {\bibfnamefont {A.}~\bibnamefont
  {Verger}}, \bibinfo {author} {\bibfnamefont {C.}~\bibnamefont {Ciuti}},\ and\
  \bibinfo {author} {\bibfnamefont {I.}~\bibnamefont {Carusotto}},\ }\href
  {https://doi.org/10.1103/PhysRevB.73.193306} {\bibfield  {journal} {\bibinfo
  {journal} {Phys. Rev. B}\ }\textbf {\bibinfo {volume} {73}},\ \bibinfo
  {pages} {193306} (\bibinfo {year} {2006})}\BibitemShut {NoStop}%
\bibitem [{\citenamefont {Mu{\~{n}}oz-Matutano}\ \emph
  {et~al.}(2019)\citenamefont {Mu{\~{n}}oz-Matutano}, \citenamefont {Wood},
  \citenamefont {Johnsson}, \citenamefont {Vidal}, \citenamefont {Baragiola},
  \citenamefont {Reinhard}, \citenamefont {Lema{\^{\i}}tre}, \citenamefont
  {Bloch}, \citenamefont {Amo}, \citenamefont {Nogues}, \citenamefont {Besga},
  \citenamefont {Richard},\ and\ \citenamefont {Volz}}]{Munoz-Matutano2019}%
  \BibitemOpen
  \bibfield  {author} {\bibinfo {author} {\bibfnamefont {G.}~\bibnamefont
  {Mu{\~{n}}oz-Matutano}}, \bibinfo {author} {\bibfnamefont {A.}~\bibnamefont
  {Wood}}, \bibinfo {author} {\bibfnamefont {M.}~\bibnamefont {Johnsson}},
  \bibinfo {author} {\bibfnamefont {X.}~\bibnamefont {Vidal}}, \bibinfo
  {author} {\bibfnamefont {B.~Q.}\ \bibnamefont {Baragiola}}, \bibinfo {author}
  {\bibfnamefont {A.}~\bibnamefont {Reinhard}}, \bibinfo {author}
  {\bibfnamefont {A.}~\bibnamefont {Lema{\^{\i}}tre}}, \bibinfo {author}
  {\bibfnamefont {J.}~\bibnamefont {Bloch}}, \bibinfo {author} {\bibfnamefont
  {A.}~\bibnamefont {Amo}}, \bibinfo {author} {\bibfnamefont {G.}~\bibnamefont
  {Nogues}}, \bibinfo {author} {\bibfnamefont {B.}~\bibnamefont {Besga}},
  \bibinfo {author} {\bibfnamefont {M.}~\bibnamefont {Richard}},\ and\ \bibinfo
  {author} {\bibfnamefont {T.}~\bibnamefont {Volz}},\ }\href
  {https://doi.org/10.1038/s41563-019-0281-z} {\bibfield  {journal} {\bibinfo
  {journal} {Nat. Mater.}\ }\textbf {\bibinfo {volume} {18}},\ \bibinfo {pages}
  {213} (\bibinfo {year} {2019})}\BibitemShut {NoStop}%
\bibitem [{\citenamefont {Delteil}\ \emph {et~al.}(2019)\citenamefont
  {Delteil}, \citenamefont {Fink}, \citenamefont {Schade}, \citenamefont
  {Höfling}, \citenamefont {Schneider},\ and\ \citenamefont
  {{\.{I}}mamo{\u{g}}lu}}]{Delteil2019}%
  \BibitemOpen
  \bibfield  {author} {\bibinfo {author} {\bibfnamefont {A.}~\bibnamefont
  {Delteil}}, \bibinfo {author} {\bibfnamefont {T.}~\bibnamefont {Fink}},
  \bibinfo {author} {\bibfnamefont {A.}~\bibnamefont {Schade}}, \bibinfo
  {author} {\bibfnamefont {S.}~\bibnamefont {Höfling}}, \bibinfo {author}
  {\bibfnamefont {C.}~\bibnamefont {Schneider}},\ and\ \bibinfo {author}
  {\bibfnamefont {A.}~\bibnamefont {{\.{I}}mamo{\u{g}}lu}},\ }\href
  {https://doi.org/10.1038/s41563-019-0282-y} {\bibfield  {journal} {\bibinfo
  {journal} {Nat. Mater.}\ }\textbf {\bibinfo {volume} {18}},\ \bibinfo {pages}
  {219} (\bibinfo {year} {2019})}\BibitemShut {NoStop}%
\bibitem [{\citenamefont {Ryou}\ \emph {et~al.}(2018)\citenamefont {Ryou},
  \citenamefont {Rosser}, \citenamefont {Saxena}, \citenamefont {Fryett},\ and\
  \citenamefont {Majumdar}}]{Ryou2018}%
  \BibitemOpen
  \bibfield  {author} {\bibinfo {author} {\bibfnamefont {A.}~\bibnamefont
  {Ryou}}, \bibinfo {author} {\bibfnamefont {D.}~\bibnamefont {Rosser}},
  \bibinfo {author} {\bibfnamefont {A.}~\bibnamefont {Saxena}}, \bibinfo
  {author} {\bibfnamefont {T.}~\bibnamefont {Fryett}},\ and\ \bibinfo {author}
  {\bibfnamefont {A.}~\bibnamefont {Majumdar}},\ }\href
  {https://doi.org/10.1103/PhysRevB.97.235307} {\bibfield  {journal} {\bibinfo
  {journal} {Phys. Rev. B}\ }\textbf {\bibinfo {volume} {97}},\ \bibinfo
  {pages} {235307} (\bibinfo {year} {2018})}\BibitemShut {NoStop}%
\bibitem [{\citenamefont {Trivedi}\ \emph {et~al.}(2019)\citenamefont
  {Trivedi}, \citenamefont {Radulaski}, \citenamefont {Fischer}, \citenamefont
  {Fan},\ and\ \citenamefont {Vu\ifmmode \check{c}\else
  \v{c}\fi{}kovi\ifmmode~\acute{c}\else \'{c}\fi{}}}]{Trivedi2019}%
  \BibitemOpen
  \bibfield  {author} {\bibinfo {author} {\bibfnamefont {R.}~\bibnamefont
  {Trivedi}}, \bibinfo {author} {\bibfnamefont {M.}~\bibnamefont {Radulaski}},
  \bibinfo {author} {\bibfnamefont {K.~A.}\ \bibnamefont {Fischer}}, \bibinfo
  {author} {\bibfnamefont {S.}~\bibnamefont {Fan}},\ and\ \bibinfo {author}
  {\bibfnamefont {J.}~\bibnamefont {Vu\ifmmode \check{c}\else
  \v{c}\fi{}kovi\ifmmode~\acute{c}\else \'{c}\fi{}}},\ }\href
  {https://doi.org/10.1103/PhysRevLett.122.243602} {\bibfield  {journal}
  {\bibinfo  {journal} {Phys. Rev. Lett.}\ }\textbf {\bibinfo {volume} {122}},\
  \bibinfo {pages} {243602} (\bibinfo {year} {2019})}\BibitemShut {NoStop}%
\bibitem [{\citenamefont {Butov}\ and\ \citenamefont
  {Kavokin}(2012)}]{Butov2012}%
  \BibitemOpen
  \bibfield  {author} {\bibinfo {author} {\bibfnamefont {L.~V.}\ \bibnamefont
  {Butov}}\ and\ \bibinfo {author} {\bibfnamefont {A.~V.}\ \bibnamefont
  {Kavokin}},\ }\href {https://doi.org/10.1038/nphoton.2011.325} {\bibfield
  {journal} {\bibinfo  {journal} {Nat. Photon.}\ }\textbf {\bibinfo {volume}
  {6}},\ \bibinfo {pages} {2} (\bibinfo {year} {2012})}\BibitemShut {NoStop}%
\bibitem [{\citenamefont {Deveaud-Pl{\'{e}}dran}(2012)}]{Deveaud-Pledran2012}%
  \BibitemOpen
  \bibfield  {author} {\bibinfo {author} {\bibfnamefont {B.}~\bibnamefont
  {Deveaud-Pl{\'{e}}dran}},\ }\href {https://doi.org/10.1038/nphoton.2012.52}
  {\bibfield  {journal} {\bibinfo  {journal} {Nat. Photon.}\ }\textbf {\bibinfo
  {volume} {6}},\ \bibinfo {pages} {205} (\bibinfo {year} {2012})}\BibitemShut
  {NoStop}%
\bibitem [{\citenamefont {Chiocchetta}\ \emph {et~al.}(2017)\citenamefont
  {Chiocchetta}, \citenamefont {Gambassi},\ and\ \citenamefont
  {Carusotto}}]{Chiocchetta2017}%
  \BibitemOpen
  \bibfield  {author} {\bibinfo {author} {\bibfnamefont {A.}~\bibnamefont
  {Chiocchetta}}, \bibinfo {author} {\bibfnamefont {A.}~\bibnamefont
  {Gambassi}},\ and\ \bibinfo {author} {\bibfnamefont {I.}~\bibnamefont
  {Carusotto}},\ }\bibinfo {title} {Laser operation and {B}ose-{E}instein
  condensation: Analogies and differences},\ in\ \href
  {https://doi.org/10.1017/9781316084366.022} {\emph {\bibinfo {booktitle}
  {Universal Themes of Bose-Einstein Condensation}}},\ \bibinfo {editor}
  {edited by\ \bibinfo {editor} {\bibfnamefont {N.~P.}\ \bibnamefont
  {Proukakis}}, \bibinfo {editor} {\bibfnamefont {D.~W.}\ \bibnamefont
  {Snoke}},\ and\ \bibinfo {editor} {\bibfnamefont {P.~B.}\ \bibnamefont
  {Littlewood}}}\ (\bibinfo  {publisher} {Cambridge University Press},\
  \bibinfo {year} {2017})\ pp.\ \bibinfo {pages} {409--423}\BibitemShut
  {NoStop}%
\bibitem [{\citenamefont {Miesner}\ \emph {et~al.}(1998)\citenamefont
  {Miesner}, \citenamefont {Stamper-Kurn}, \citenamefont {Andrews},
  \citenamefont {Durfee}, \citenamefont {Inouye},\ and\ \citenamefont
  {Ketterle}}]{Miesner1998}%
  \BibitemOpen
  \bibfield  {author} {\bibinfo {author} {\bibfnamefont {H.-J.}\ \bibnamefont
  {Miesner}}, \bibinfo {author} {\bibfnamefont {D.~M.}\ \bibnamefont
  {Stamper-Kurn}}, \bibinfo {author} {\bibfnamefont {M.~R.}\ \bibnamefont
  {Andrews}}, \bibinfo {author} {\bibfnamefont {D.~S.}\ \bibnamefont {Durfee}},
  \bibinfo {author} {\bibfnamefont {S.}~\bibnamefont {Inouye}},\ and\ \bibinfo
  {author} {\bibfnamefont {W.}~\bibnamefont {Ketterle}},\ }\href
  {https://doi.org/10.1126/science.279.5353.1005} {\bibfield  {journal}
  {\bibinfo  {journal} {Science}\ }\textbf {\bibinfo {volume} {279}},\ \bibinfo
  {pages} {1005} (\bibinfo {year} {1998})}\BibitemShut {NoStop}%
\bibitem [{\citenamefont {Ketterle}\ and\ \citenamefont
  {Inouye}(2001)}]{Ketterle2001}%
  \BibitemOpen
  \bibfield  {author} {\bibinfo {author} {\bibfnamefont {W.}~\bibnamefont
  {Ketterle}}\ and\ \bibinfo {author} {\bibfnamefont {S.}~\bibnamefont
  {Inouye}},\ }\href {https://doi.org/10.1016/S1296-2147(01)01180-5} {\bibfield
   {journal} {\bibinfo  {journal} {Comptes Rendus Acad. Sci.}\ }\textbf
  {\bibinfo {volume} {2}},\ \bibinfo {pages} {339 } (\bibinfo {year}
  {2001})}\BibitemShut {NoStop}%
\bibitem [{\citenamefont {Snoke}(2012)}]{Snoke2012}%
  \BibitemOpen
  \bibfield  {author} {\bibinfo {author} {\bibfnamefont {D.}~\bibnamefont
  {Snoke}},\ }\bibinfo {title} {Polariton condensation and lasing},\ in\ \href
  {https://doi.org/10.1007/978-3-642-24186-4_12} {\emph {\bibinfo {booktitle}
  {Exciton Polaritons in Microcavities: New Frontiers}}},\ \bibinfo {editor}
  {edited by\ \bibinfo {editor} {\bibfnamefont {V.}~\bibnamefont {Timofeev}}\
  and\ \bibinfo {editor} {\bibfnamefont {D.}~\bibnamefont {Sanvitto}}}\
  (\bibinfo  {publisher} {Springer},\ \bibinfo {year} {2012})\ pp.\ \bibinfo
  {pages} {307--327}\BibitemShut {NoStop}%
\bibitem [{\citenamefont {Imamoglu}\ \emph {et~al.}(1996)\citenamefont
  {Imamoglu}, \citenamefont {Ram}, \citenamefont {Pau},\ and\ \citenamefont
  {Yamamoto}}]{Imamoglu1996}%
  \BibitemOpen
  \bibfield  {author} {\bibinfo {author} {\bibfnamefont {A.}~\bibnamefont
  {Imamoglu}}, \bibinfo {author} {\bibfnamefont {R.~J.}\ \bibnamefont {Ram}},
  \bibinfo {author} {\bibfnamefont {S.}~\bibnamefont {Pau}},\ and\ \bibinfo
  {author} {\bibfnamefont {Y.}~\bibnamefont {Yamamoto}},\ }\href
  {https://doi.org/10.1103/PhysRevA.53.4250} {\bibfield  {journal} {\bibinfo
  {journal} {Phys. Rev. A}\ }\textbf {\bibinfo {volume} {53}},\ \bibinfo
  {pages} {4250} (\bibinfo {year} {1996})}\BibitemShut {NoStop}%
\bibitem [{\citenamefont {Deng}\ \emph {et~al.}(2003)\citenamefont {Deng},
  \citenamefont {Weihs}, \citenamefont {Snoke}, \citenamefont {Bloch},\ and\
  \citenamefont {Yamamoto}}]{Deng2003}%
  \BibitemOpen
  \bibfield  {author} {\bibinfo {author} {\bibfnamefont {H.}~\bibnamefont
  {Deng}}, \bibinfo {author} {\bibfnamefont {G.}~\bibnamefont {Weihs}},
  \bibinfo {author} {\bibfnamefont {D.}~\bibnamefont {Snoke}}, \bibinfo
  {author} {\bibfnamefont {J.}~\bibnamefont {Bloch}},\ and\ \bibinfo {author}
  {\bibfnamefont {Y.}~\bibnamefont {Yamamoto}},\ }\href
  {https://doi.org/10.1073/pnas.2634328100} {\bibfield  {journal} {\bibinfo
  {journal} {Proc. Natl. Acad. Sci. U.S.A.}\ }\textbf {\bibinfo {volume}
  {100}},\ \bibinfo {pages} {15318} (\bibinfo {year} {2003})}\BibitemShut
  {NoStop}%
\bibitem [{\citenamefont {Deng}\ \emph {et~al.}(2006)\citenamefont {Deng},
  \citenamefont {Press}, \citenamefont {G\"otzinger}, \citenamefont {Solomon},
  \citenamefont {Hey}, \citenamefont {Ploog},\ and\ \citenamefont
  {Yamamoto}}]{Deng2006}%
  \BibitemOpen
  \bibfield  {author} {\bibinfo {author} {\bibfnamefont {H.}~\bibnamefont
  {Deng}}, \bibinfo {author} {\bibfnamefont {D.}~\bibnamefont {Press}},
  \bibinfo {author} {\bibfnamefont {S.}~\bibnamefont {G\"otzinger}}, \bibinfo
  {author} {\bibfnamefont {G.~S.}\ \bibnamefont {Solomon}}, \bibinfo {author}
  {\bibfnamefont {R.}~\bibnamefont {Hey}}, \bibinfo {author} {\bibfnamefont
  {K.~H.}\ \bibnamefont {Ploog}},\ and\ \bibinfo {author} {\bibfnamefont
  {Y.}~\bibnamefont {Yamamoto}},\ }\href
  {https://doi.org/10.1103/PhysRevLett.97.146402} {\bibfield  {journal}
  {\bibinfo  {journal} {Phys. Rev. Lett.}\ }\textbf {\bibinfo {volume} {97}},\
  \bibinfo {pages} {146402} (\bibinfo {year} {2006})}\BibitemShut {NoStop}%
\bibitem [{\citenamefont {Kasprzak}\ \emph {et~al.}(2008)\citenamefont
  {Kasprzak}, \citenamefont {Solnyshkov}, \citenamefont {Andr\'e},
  \citenamefont {Dang},\ and\ \citenamefont {Malpuech}}]{Kasprzak2008}%
  \BibitemOpen
  \bibfield  {author} {\bibinfo {author} {\bibfnamefont {J.}~\bibnamefont
  {Kasprzak}}, \bibinfo {author} {\bibfnamefont {D.~D.}\ \bibnamefont
  {Solnyshkov}}, \bibinfo {author} {\bibfnamefont {R.}~\bibnamefont {Andr\'e}},
  \bibinfo {author} {\bibfnamefont {L.~S.}\ \bibnamefont {Dang}},\ and\
  \bibinfo {author} {\bibfnamefont {G.}~\bibnamefont {Malpuech}},\ }\href
  {https://doi.org/10.1103/PhysRevLett.101.146404} {\bibfield  {journal}
  {\bibinfo  {journal} {Phys. Rev. Lett.}\ }\textbf {\bibinfo {volume} {101}},\
  \bibinfo {pages} {146404} (\bibinfo {year} {2008})}\BibitemShut {NoStop}%
\bibitem [{\citenamefont {Byrnes}\ \emph {et~al.}(2014)\citenamefont {Byrnes},
  \citenamefont {Kim},\ and\ \citenamefont {Yamamoto}}]{Byrnes2014}%
  \BibitemOpen
  \bibfield  {author} {\bibinfo {author} {\bibfnamefont {T.}~\bibnamefont
  {Byrnes}}, \bibinfo {author} {\bibfnamefont {N.~Y.}\ \bibnamefont {Kim}},\
  and\ \bibinfo {author} {\bibfnamefont {Y.}~\bibnamefont {Yamamoto}},\ }\href
  {https://doi.org/10.1038/nphys3143} {\bibfield  {journal} {\bibinfo
  {journal} {Nat. Phys.}\ }\textbf {\bibinfo {volume} {10}},\ \bibinfo {pages}
  {803} (\bibinfo {year} {2014})}\BibitemShut {NoStop}%
\bibitem [{\citenamefont {Sun}\ \emph {et~al.}(2017{\natexlab{a}})\citenamefont
  {Sun}, \citenamefont {Wen}, \citenamefont {Yoon}, \citenamefont {Liu},
  \citenamefont {Steger}, \citenamefont {Pfeiffer}, \citenamefont {West},
  \citenamefont {Snoke},\ and\ \citenamefont {Nelson}}]{Sun2017}%
  \BibitemOpen
  \bibfield  {author} {\bibinfo {author} {\bibfnamefont {Y.}~\bibnamefont
  {Sun}}, \bibinfo {author} {\bibfnamefont {P.}~\bibnamefont {Wen}}, \bibinfo
  {author} {\bibfnamefont {Y.}~\bibnamefont {Yoon}}, \bibinfo {author}
  {\bibfnamefont {G.}~\bibnamefont {Liu}}, \bibinfo {author} {\bibfnamefont
  {M.}~\bibnamefont {Steger}}, \bibinfo {author} {\bibfnamefont {L.~N.}\
  \bibnamefont {Pfeiffer}}, \bibinfo {author} {\bibfnamefont {K.}~\bibnamefont
  {West}}, \bibinfo {author} {\bibfnamefont {D.~W.}\ \bibnamefont {Snoke}},\
  and\ \bibinfo {author} {\bibfnamefont {K.~A.}\ \bibnamefont {Nelson}},\
  }\href {https://doi.org/10.1103/PhysRevLett.118.016602} {\bibfield  {journal}
  {\bibinfo  {journal} {Phys. Rev. Lett.}\ }\textbf {\bibinfo {volume} {118}},\
  \bibinfo {pages} {016602} (\bibinfo {year} {2017}{\natexlab{a}})}\BibitemShut
  {NoStop}%
\bibitem [{\citenamefont {Caputo}\ \emph {et~al.}(2017)\citenamefont {Caputo},
  \citenamefont {Ballarini}, \citenamefont {Dagvadorj}, \citenamefont
  {Mu{\~{n}}oz}, \citenamefont {Giorgi}, \citenamefont {Dominici},
  \citenamefont {West}, \citenamefont {Pfeiffer}, \citenamefont {Gigli},
  \citenamefont {Laussy}, \citenamefont {Szyma{\'{n}}ska},\ and\ \citenamefont
  {Sanvitto}}]{Caputo2017}%
  \BibitemOpen
  \bibfield  {author} {\bibinfo {author} {\bibfnamefont {D.}~\bibnamefont
  {Caputo}}, \bibinfo {author} {\bibfnamefont {D.}~\bibnamefont {Ballarini}},
  \bibinfo {author} {\bibfnamefont {G.}~\bibnamefont {Dagvadorj}}, \bibinfo
  {author} {\bibfnamefont {C.~S.}\ \bibnamefont {Mu{\~{n}}oz}}, \bibinfo
  {author} {\bibfnamefont {M.~D.}\ \bibnamefont {Giorgi}}, \bibinfo {author}
  {\bibfnamefont {L.}~\bibnamefont {Dominici}}, \bibinfo {author}
  {\bibfnamefont {K.}~\bibnamefont {West}}, \bibinfo {author} {\bibfnamefont
  {L.~N.}\ \bibnamefont {Pfeiffer}}, \bibinfo {author} {\bibfnamefont
  {G.}~\bibnamefont {Gigli}}, \bibinfo {author} {\bibfnamefont {F.~P.}\
  \bibnamefont {Laussy}}, \bibinfo {author} {\bibfnamefont {M.~H.}\
  \bibnamefont {Szyma{\'{n}}ska}},\ and\ \bibinfo {author} {\bibfnamefont
  {D.}~\bibnamefont {Sanvitto}},\ }\href {https://doi.org/10.1038/nmat5039}
  {\bibfield  {journal} {\bibinfo  {journal} {Nat. Mater.}\ }\textbf {\bibinfo
  {volume} {17}},\ \bibinfo {pages} {145} (\bibinfo {year} {2017})}\BibitemShut
  {NoStop}%
\bibitem [{\citenamefont {Klaers}\ \emph
  {et~al.}(2010{\natexlab{a}})\citenamefont {Klaers}, \citenamefont
  {Vewinger},\ and\ \citenamefont {Weitz}}]{Klaers2010a}%
  \BibitemOpen
  \bibfield  {author} {\bibinfo {author} {\bibfnamefont {J.}~\bibnamefont
  {Klaers}}, \bibinfo {author} {\bibfnamefont {F.}~\bibnamefont {Vewinger}},\
  and\ \bibinfo {author} {\bibfnamefont {M.}~\bibnamefont {Weitz}},\ }\href
  {https://doi.org/10.1038/nphys1680} {\bibfield  {journal} {\bibinfo
  {journal} {Nat. Phys.}\ }\textbf {\bibinfo {volume} {6}},\ \bibinfo {pages}
  {512} (\bibinfo {year} {2010}{\natexlab{a}})}\BibitemShut {NoStop}%
\bibitem [{\citenamefont {Klaers}\ \emph
  {et~al.}(2010{\natexlab{b}})\citenamefont {Klaers}, \citenamefont {Schmitt},
  \citenamefont {Vewinger},\ and\ \citenamefont {Weitz}}]{Klaers2010}%
  \BibitemOpen
  \bibfield  {author} {\bibinfo {author} {\bibfnamefont {J.}~\bibnamefont
  {Klaers}}, \bibinfo {author} {\bibfnamefont {J.}~\bibnamefont {Schmitt}},
  \bibinfo {author} {\bibfnamefont {F.}~\bibnamefont {Vewinger}},\ and\
  \bibinfo {author} {\bibfnamefont {M.}~\bibnamefont {Weitz}},\ }\href
  {https://doi.org/10.1038/nature09567} {\bibfield  {journal} {\bibinfo
  {journal} {Nature}\ }\textbf {\bibinfo {volume} {468}},\ \bibinfo {pages}
  {545} (\bibinfo {year} {2010}{\natexlab{b}})}\BibitemShut {NoStop}%
\bibitem [{\citenamefont {Schneider}\ \emph {et~al.}(2020)\citenamefont
  {Schneider}, \citenamefont {Br\"{a}cher}, \citenamefont {Breitbach},
  \citenamefont {Lauer}, \citenamefont {Pirro}, \citenamefont {Bozhko},
  \citenamefont {Musiienko-Shmarova}, \citenamefont {Heinz}, \citenamefont
  {Wang}, \citenamefont {Meyer}, \citenamefont {Heussner}, \citenamefont
  {Keller}, \citenamefont {Papaioannou}, \citenamefont {L\"{a}gel},
  \citenamefont {L\"{o}ber}, \citenamefont {Dubs}, \citenamefont {Slavin},
  \citenamefont {Tiberkevich}, \citenamefont {Serga}, \citenamefont
  {Hillebrands},\ and\ \citenamefont {Chumak}}]{Schneider2020}%
  \BibitemOpen
  \bibfield  {author} {\bibinfo {author} {\bibfnamefont {M.}~\bibnamefont
  {Schneider}}, \bibinfo {author} {\bibfnamefont {T.}~\bibnamefont
  {Br\"{a}cher}}, \bibinfo {author} {\bibfnamefont {D.}~\bibnamefont
  {Breitbach}}, \bibinfo {author} {\bibfnamefont {V.}~\bibnamefont {Lauer}},
  \bibinfo {author} {\bibfnamefont {P.}~\bibnamefont {Pirro}}, \bibinfo
  {author} {\bibfnamefont {D.~A.}\ \bibnamefont {Bozhko}}, \bibinfo {author}
  {\bibfnamefont {H.~Y.}\ \bibnamefont {Musiienko-Shmarova}}, \bibinfo {author}
  {\bibfnamefont {B.}~\bibnamefont {Heinz}}, \bibinfo {author} {\bibfnamefont
  {Q.}~\bibnamefont {Wang}}, \bibinfo {author} {\bibfnamefont {T.}~\bibnamefont
  {Meyer}}, \bibinfo {author} {\bibfnamefont {F.}~\bibnamefont {Heussner}},
  \bibinfo {author} {\bibfnamefont {S.}~\bibnamefont {Keller}}, \bibinfo
  {author} {\bibfnamefont {E.~T.}\ \bibnamefont {Papaioannou}}, \bibinfo
  {author} {\bibfnamefont {B.}~\bibnamefont {L\"{a}gel}}, \bibinfo {author}
  {\bibfnamefont {T.}~\bibnamefont {L\"{o}ber}}, \bibinfo {author}
  {\bibfnamefont {C.}~\bibnamefont {Dubs}}, \bibinfo {author} {\bibfnamefont
  {A.~N.}\ \bibnamefont {Slavin}}, \bibinfo {author} {\bibfnamefont {V.~S.}\
  \bibnamefont {Tiberkevich}}, \bibinfo {author} {\bibfnamefont {A.~A.}\
  \bibnamefont {Serga}}, \bibinfo {author} {\bibfnamefont {B.}~\bibnamefont
  {Hillebrands}},\ and\ \bibinfo {author} {\bibfnamefont {A.~V.}\ \bibnamefont
  {Chumak}},\ }\href {https://doi.org/10.1038/s41565-020-0671-z} {\bibfield
  {journal} {\bibinfo  {journal} {Nat. Nanotechnol.}\ }\textbf {\bibinfo
  {volume} {15}},\ \bibinfo {pages} {457} (\bibinfo {year} {2020})}\BibitemShut
  {NoStop}%
\bibitem [{\citenamefont {Schneider}\ \emph {et~al.}(2016)\citenamefont
  {Schneider}, \citenamefont {Winkler}, \citenamefont {Fraser}, \citenamefont
  {Kamp}, \citenamefont {Yamamoto}, \citenamefont {Ostrovskaya},\ and\
  \citenamefont {Höfling}}]{Schneider2016}%
  \BibitemOpen
  \bibfield  {author} {\bibinfo {author} {\bibfnamefont {C.}~\bibnamefont
  {Schneider}}, \bibinfo {author} {\bibfnamefont {K.}~\bibnamefont {Winkler}},
  \bibinfo {author} {\bibfnamefont {M.~D.}\ \bibnamefont {Fraser}}, \bibinfo
  {author} {\bibfnamefont {M.}~\bibnamefont {Kamp}}, \bibinfo {author}
  {\bibfnamefont {Y.}~\bibnamefont {Yamamoto}}, \bibinfo {author}
  {\bibfnamefont {E.~A.}\ \bibnamefont {Ostrovskaya}},\ and\ \bibinfo {author}
  {\bibfnamefont {S.}~\bibnamefont {Höfling}},\ }\href
  {https://doi.org/10.1088/0034-4885/80/1/016503} {\bibfield  {journal}
  {\bibinfo  {journal} {Rep. Prog. Phys.}\ }\textbf {\bibinfo {volume} {80}},\
  \bibinfo {pages} {016503} (\bibinfo {year} {2016})}\BibitemShut {NoStop}%
\bibitem [{\citenamefont {Cristofolini}\ \emph {et~al.}(2018)\citenamefont
  {Cristofolini}, \citenamefont {Hatzopoulos}, \citenamefont {Savvidis},\ and\
  \citenamefont {Baumberg}}]{Cristofolini2018}%
  \BibitemOpen
  \bibfield  {author} {\bibinfo {author} {\bibfnamefont {P.}~\bibnamefont
  {Cristofolini}}, \bibinfo {author} {\bibfnamefont {Z.}~\bibnamefont
  {Hatzopoulos}}, \bibinfo {author} {\bibfnamefont {P.~G.}\ \bibnamefont
  {Savvidis}},\ and\ \bibinfo {author} {\bibfnamefont {J.~J.}\ \bibnamefont
  {Baumberg}},\ }\href {https://doi.org/10.1103/PhysRevLett.121.067401}
  {\bibfield  {journal} {\bibinfo  {journal} {Phys. Rev. Lett.}\ }\textbf
  {\bibinfo {volume} {121}},\ \bibinfo {pages} {067401} (\bibinfo {year}
  {2018})}\BibitemShut {NoStop}%
\bibitem [{\citenamefont {Ballarini}\ \emph {et~al.}(2019)\citenamefont
  {Ballarini}, \citenamefont {Chestnov}, \citenamefont {Caputo}, \citenamefont
  {De~Giorgi}, \citenamefont {Dominici}, \citenamefont {West}, \citenamefont
  {Pfeiffer}, \citenamefont {Gigli}, \citenamefont {Kavokin},\ and\
  \citenamefont {Sanvitto}}]{Ballarini2019}%
  \BibitemOpen
  \bibfield  {author} {\bibinfo {author} {\bibfnamefont {D.}~\bibnamefont
  {Ballarini}}, \bibinfo {author} {\bibfnamefont {I.}~\bibnamefont {Chestnov}},
  \bibinfo {author} {\bibfnamefont {D.}~\bibnamefont {Caputo}}, \bibinfo
  {author} {\bibfnamefont {M.}~\bibnamefont {De~Giorgi}}, \bibinfo {author}
  {\bibfnamefont {L.}~\bibnamefont {Dominici}}, \bibinfo {author}
  {\bibfnamefont {K.}~\bibnamefont {West}}, \bibinfo {author} {\bibfnamefont
  {L.~N.}\ \bibnamefont {Pfeiffer}}, \bibinfo {author} {\bibfnamefont
  {G.}~\bibnamefont {Gigli}}, \bibinfo {author} {\bibfnamefont
  {A.}~\bibnamefont {Kavokin}},\ and\ \bibinfo {author} {\bibfnamefont
  {D.}~\bibnamefont {Sanvitto}},\ }\href
  {https://doi.org/10.1103/PhysRevLett.123.047401} {\bibfield  {journal}
  {\bibinfo  {journal} {Phys. Rev. Lett.}\ }\textbf {\bibinfo {volume} {123}},\
  \bibinfo {pages} {047401} (\bibinfo {year} {2019})}\BibitemShut {NoStop}%
\bibitem [{\citenamefont {Wertz}\ \emph {et~al.}(2010)\citenamefont {Wertz},
  \citenamefont {Ferrier}, \citenamefont {Solnyshkov}, \citenamefont {Johne},
  \citenamefont {Sanvitto}, \citenamefont {Lema{\^{\i}}tre}, \citenamefont
  {Sagnes}, \citenamefont {Grousson}, \citenamefont {Kavokin}, \citenamefont
  {Senellart}, \citenamefont {Malpuech},\ and\ \citenamefont
  {Bloch}}]{Wertz2010}%
  \BibitemOpen
  \bibfield  {author} {\bibinfo {author} {\bibfnamefont {E.}~\bibnamefont
  {Wertz}}, \bibinfo {author} {\bibfnamefont {L.}~\bibnamefont {Ferrier}},
  \bibinfo {author} {\bibfnamefont {D.~D.}\ \bibnamefont {Solnyshkov}},
  \bibinfo {author} {\bibfnamefont {R.}~\bibnamefont {Johne}}, \bibinfo
  {author} {\bibfnamefont {D.}~\bibnamefont {Sanvitto}}, \bibinfo {author}
  {\bibfnamefont {A.}~\bibnamefont {Lema{\^{\i}}tre}}, \bibinfo {author}
  {\bibfnamefont {I.}~\bibnamefont {Sagnes}}, \bibinfo {author} {\bibfnamefont
  {R.}~\bibnamefont {Grousson}}, \bibinfo {author} {\bibfnamefont {A.~V.}\
  \bibnamefont {Kavokin}}, \bibinfo {author} {\bibfnamefont {P.}~\bibnamefont
  {Senellart}}, \bibinfo {author} {\bibfnamefont {G.}~\bibnamefont
  {Malpuech}},\ and\ \bibinfo {author} {\bibfnamefont {J.}~\bibnamefont
  {Bloch}},\ }\href {https://doi.org/10.1038/nphys1750} {\bibfield  {journal}
  {\bibinfo  {journal} {Nat. Phys.}\ }\textbf {\bibinfo {volume} {6}},\
  \bibinfo {pages} {860} (\bibinfo {year} {2010})}\BibitemShut {NoStop}%
\bibitem [{\citenamefont {Christmann}\ \emph {et~al.}(2012)\citenamefont
  {Christmann}, \citenamefont {Tosi}, \citenamefont {Berloff}, \citenamefont
  {Tsotsis}, \citenamefont {Eldridge}, \citenamefont {Hatzopoulos},
  \citenamefont {Savvidis},\ and\ \citenamefont {Baumberg}}]{Christmann2012}%
  \BibitemOpen
  \bibfield  {author} {\bibinfo {author} {\bibfnamefont {G.}~\bibnamefont
  {Christmann}}, \bibinfo {author} {\bibfnamefont {G.}~\bibnamefont {Tosi}},
  \bibinfo {author} {\bibfnamefont {N.~G.}\ \bibnamefont {Berloff}}, \bibinfo
  {author} {\bibfnamefont {P.}~\bibnamefont {Tsotsis}}, \bibinfo {author}
  {\bibfnamefont {P.~S.}\ \bibnamefont {Eldridge}}, \bibinfo {author}
  {\bibfnamefont {Z.}~\bibnamefont {Hatzopoulos}}, \bibinfo {author}
  {\bibfnamefont {P.~G.}\ \bibnamefont {Savvidis}},\ and\ \bibinfo {author}
  {\bibfnamefont {J.~J.}\ \bibnamefont {Baumberg}},\ }\href
  {https://doi.org/10.1103/PhysRevB.85.235303} {\bibfield  {journal} {\bibinfo
  {journal} {Phys. Rev. B}\ }\textbf {\bibinfo {volume} {85}},\ \bibinfo
  {pages} {235303} (\bibinfo {year} {2012})}\BibitemShut {NoStop}%
\bibitem [{\citenamefont {Steger}\ \emph {et~al.}(2015)\citenamefont {Steger},
  \citenamefont {Gautham}, \citenamefont {Snoke}, \citenamefont {Pfeiffer},\
  and\ \citenamefont {West}}]{Steger2015}%
  \BibitemOpen
  \bibfield  {author} {\bibinfo {author} {\bibfnamefont {M.}~\bibnamefont
  {Steger}}, \bibinfo {author} {\bibfnamefont {C.}~\bibnamefont {Gautham}},
  \bibinfo {author} {\bibfnamefont {D.~W.}\ \bibnamefont {Snoke}}, \bibinfo
  {author} {\bibfnamefont {L.}~\bibnamefont {Pfeiffer}},\ and\ \bibinfo
  {author} {\bibfnamefont {K.}~\bibnamefont {West}},\ }\href
  {https://doi.org/10.1364/optica.2.000001} {\bibfield  {journal} {\bibinfo
  {journal} {Optica}\ }\textbf {\bibinfo {volume} {2}},\ \bibinfo {pages} {1}
  (\bibinfo {year} {2015})}\BibitemShut {NoStop}%
\bibitem [{\citenamefont {Mukherjee}\ \emph {et~al.}(2019)\citenamefont
  {Mukherjee}, \citenamefont {Myers}, \citenamefont {Lena}, \citenamefont
  {Ozden}, \citenamefont {Beaumariage}, \citenamefont {Sun}, \citenamefont
  {Steger}, \citenamefont {Pfeiffer}, \citenamefont {West}, \citenamefont
  {Daley},\ and\ \citenamefont {Snoke}}]{Mukherjee2019}%
  \BibitemOpen
  \bibfield  {author} {\bibinfo {author} {\bibfnamefont {S.}~\bibnamefont
  {Mukherjee}}, \bibinfo {author} {\bibfnamefont {D.~M.}\ \bibnamefont
  {Myers}}, \bibinfo {author} {\bibfnamefont {R.~G.}\ \bibnamefont {Lena}},
  \bibinfo {author} {\bibfnamefont {B.}~\bibnamefont {Ozden}}, \bibinfo
  {author} {\bibfnamefont {J.}~\bibnamefont {Beaumariage}}, \bibinfo {author}
  {\bibfnamefont {Z.}~\bibnamefont {Sun}}, \bibinfo {author} {\bibfnamefont
  {M.}~\bibnamefont {Steger}}, \bibinfo {author} {\bibfnamefont {L.~N.}\
  \bibnamefont {Pfeiffer}}, \bibinfo {author} {\bibfnamefont {K.}~\bibnamefont
  {West}}, \bibinfo {author} {\bibfnamefont {A.~J.}\ \bibnamefont {Daley}},\
  and\ \bibinfo {author} {\bibfnamefont {D.~W.}\ \bibnamefont {Snoke}},\ }\href
  {https://doi.org/10.1103/PhysRevB.100.245304} {\bibfield  {journal} {\bibinfo
   {journal} {Phys. Rev. B}\ }\textbf {\bibinfo {volume} {100}},\ \bibinfo
  {pages} {245304} (\bibinfo {year} {2019})}\BibitemShut {NoStop}%
\bibitem [{\citenamefont {Colas}\ and\ \citenamefont
  {Laussy}(2016)}]{Colas2016}%
  \BibitemOpen
  \bibfield  {author} {\bibinfo {author} {\bibfnamefont {D.}~\bibnamefont
  {Colas}}\ and\ \bibinfo {author} {\bibfnamefont {F.~P.}\ \bibnamefont
  {Laussy}},\ }\href {https://doi.org/10.1103/PhysRevLett.116.026401}
  {\bibfield  {journal} {\bibinfo  {journal} {Phys. Rev. Lett.}\ }\textbf
  {\bibinfo {volume} {116}},\ \bibinfo {pages} {026401} (\bibinfo {year}
  {2016})}\BibitemShut {NoStop}%
\bibitem [{\citenamefont {Khamehchi}\ \emph {et~al.}(2017)\citenamefont
  {Khamehchi}, \citenamefont {Hossain}, \citenamefont {Mossman}, \citenamefont
  {Zhang}, \citenamefont {Busch}, \citenamefont {Forbes},\ and\ \citenamefont
  {Engels}}]{Khamehchi2017}%
  \BibitemOpen
  \bibfield  {author} {\bibinfo {author} {\bibfnamefont {M.~A.}\ \bibnamefont
  {Khamehchi}}, \bibinfo {author} {\bibfnamefont {K.}~\bibnamefont {Hossain}},
  \bibinfo {author} {\bibfnamefont {M.~E.}\ \bibnamefont {Mossman}}, \bibinfo
  {author} {\bibfnamefont {Y.}~\bibnamefont {Zhang}}, \bibinfo {author}
  {\bibfnamefont {T.}~\bibnamefont {Busch}}, \bibinfo {author} {\bibfnamefont
  {M.~M.}\ \bibnamefont {Forbes}},\ and\ \bibinfo {author} {\bibfnamefont
  {P.}~\bibnamefont {Engels}},\ }\href
  {https://doi.org/10.1103/PhysRevLett.118.155301} {\bibfield  {journal}
  {\bibinfo  {journal} {Phys. Rev. Lett.}\ }\textbf {\bibinfo {volume} {118}},\
  \bibinfo {pages} {155301} (\bibinfo {year} {2017})}\BibitemShut {NoStop}%
\bibitem [{\citenamefont {Myers}\ \emph {et~al.}(2018)\citenamefont {Myers},
  \citenamefont {Mukherjee}, \citenamefont {Beaumariage}, \citenamefont
  {Snoke}, \citenamefont {Steger}, \citenamefont {Pfeiffer},\ and\
  \citenamefont {West}}]{Myers2018}%
  \BibitemOpen
  \bibfield  {author} {\bibinfo {author} {\bibfnamefont {D.~M.}\ \bibnamefont
  {Myers}}, \bibinfo {author} {\bibfnamefont {S.}~\bibnamefont {Mukherjee}},
  \bibinfo {author} {\bibfnamefont {J.}~\bibnamefont {Beaumariage}}, \bibinfo
  {author} {\bibfnamefont {D.~W.}\ \bibnamefont {Snoke}}, \bibinfo {author}
  {\bibfnamefont {M.}~\bibnamefont {Steger}}, \bibinfo {author} {\bibfnamefont
  {L.~N.}\ \bibnamefont {Pfeiffer}},\ and\ \bibinfo {author} {\bibfnamefont
  {K.}~\bibnamefont {West}},\ }\href
  {https://doi.org/10.1103/PhysRevB.98.235302} {\bibfield  {journal} {\bibinfo
  {journal} {Phys. Rev. B}\ }\textbf {\bibinfo {volume} {98}},\ \bibinfo
  {pages} {235302} (\bibinfo {year} {2018})}\BibitemShut {NoStop}%
\bibitem [{\citenamefont {Wouters}\ \emph {et~al.}(2010)\citenamefont
  {Wouters}, \citenamefont {Liew},\ and\ \citenamefont {Savona}}]{Wouters2010}%
  \BibitemOpen
  \bibfield  {author} {\bibinfo {author} {\bibfnamefont {M.}~\bibnamefont
  {Wouters}}, \bibinfo {author} {\bibfnamefont {T.~C.~H.}\ \bibnamefont
  {Liew}},\ and\ \bibinfo {author} {\bibfnamefont {V.}~\bibnamefont {Savona}},\
  }\href {https://doi.org/10.1103/PhysRevB.82.245315} {\bibfield  {journal}
  {\bibinfo  {journal} {Phys. Rev. B}\ }\textbf {\bibinfo {volume} {82}},\
  \bibinfo {pages} {245315} (\bibinfo {year} {2010})}\BibitemShut {NoStop}%
\bibitem [{\citenamefont {Ballarini}\ \emph {et~al.}(2017)\citenamefont
  {Ballarini}, \citenamefont {Caputo}, \citenamefont {Mu\~noz}, \citenamefont
  {De~Giorgi}, \citenamefont {Dominici}, \citenamefont
  {Szyma\ifmmode~\acute{n}\else \'{n}\fi{}ska}, \citenamefont {West},
  \citenamefont {Pfeiffer}, \citenamefont {Gigli}, \citenamefont {Laussy},\
  and\ \citenamefont {Sanvitto}}]{Ballarini2017}%
  \BibitemOpen
  \bibfield  {author} {\bibinfo {author} {\bibfnamefont {D.}~\bibnamefont
  {Ballarini}}, \bibinfo {author} {\bibfnamefont {D.}~\bibnamefont {Caputo}},
  \bibinfo {author} {\bibfnamefont {C.~S.}\ \bibnamefont {Mu\~noz}}, \bibinfo
  {author} {\bibfnamefont {M.}~\bibnamefont {De~Giorgi}}, \bibinfo {author}
  {\bibfnamefont {L.}~\bibnamefont {Dominici}}, \bibinfo {author}
  {\bibfnamefont {M.~H.}\ \bibnamefont {Szyma\ifmmode~\acute{n}\else
  \'{n}\fi{}ska}}, \bibinfo {author} {\bibfnamefont {K.}~\bibnamefont {West}},
  \bibinfo {author} {\bibfnamefont {L.~N.}\ \bibnamefont {Pfeiffer}}, \bibinfo
  {author} {\bibfnamefont {G.}~\bibnamefont {Gigli}}, \bibinfo {author}
  {\bibfnamefont {F.~P.}\ \bibnamefont {Laussy}},\ and\ \bibinfo {author}
  {\bibfnamefont {D.}~\bibnamefont {Sanvitto}},\ }\href
  {https://doi.org/10.1103/PhysRevLett.118.215301} {\bibfield  {journal}
  {\bibinfo  {journal} {Phys. Rev. Lett.}\ }\textbf {\bibinfo {volume} {118}},\
  \bibinfo {pages} {215301} (\bibinfo {year} {2017})}\BibitemShut {NoStop}%
\bibitem [{\citenamefont {Rivera}\ \emph {et~al.}(2016)\citenamefont {Rivera},
  \citenamefont {Seyler}, \citenamefont {Yu}, \citenamefont {Schaibley},
  \citenamefont {Yan}, \citenamefont {Mandrus}, \citenamefont {Yao},\ and\
  \citenamefont {Xu}}]{Rivera2016}%
  \BibitemOpen
  \bibfield  {author} {\bibinfo {author} {\bibfnamefont {P.}~\bibnamefont
  {Rivera}}, \bibinfo {author} {\bibfnamefont {K.~L.}\ \bibnamefont {Seyler}},
  \bibinfo {author} {\bibfnamefont {H.}~\bibnamefont {Yu}}, \bibinfo {author}
  {\bibfnamefont {J.~R.}\ \bibnamefont {Schaibley}}, \bibinfo {author}
  {\bibfnamefont {J.}~\bibnamefont {Yan}}, \bibinfo {author} {\bibfnamefont
  {D.~G.}\ \bibnamefont {Mandrus}}, \bibinfo {author} {\bibfnamefont
  {W.}~\bibnamefont {Yao}},\ and\ \bibinfo {author} {\bibfnamefont
  {X.}~\bibnamefont {Xu}},\ }\href {https://doi.org/10.1126/science.aac7820}
  {\bibfield  {journal} {\bibinfo  {journal} {Science}\ }\textbf {\bibinfo
  {volume} {351}},\ \bibinfo {pages} {688} (\bibinfo {year}
  {2016})}\BibitemShut {NoStop}%
\bibitem [{\citenamefont {Kulig}\ \emph {et~al.}(2018)\citenamefont {Kulig},
  \citenamefont {Zipfel}, \citenamefont {Nagler}, \citenamefont {Blanter},
  \citenamefont {Sch\"uller}, \citenamefont {Korn}, \citenamefont {Paradiso},
  \citenamefont {Glazov},\ and\ \citenamefont {Chernikov}}]{Kulig2018}%
  \BibitemOpen
  \bibfield  {author} {\bibinfo {author} {\bibfnamefont {M.}~\bibnamefont
  {Kulig}}, \bibinfo {author} {\bibfnamefont {J.}~\bibnamefont {Zipfel}},
  \bibinfo {author} {\bibfnamefont {P.}~\bibnamefont {Nagler}}, \bibinfo
  {author} {\bibfnamefont {S.}~\bibnamefont {Blanter}}, \bibinfo {author}
  {\bibfnamefont {C.}~\bibnamefont {Sch\"uller}}, \bibinfo {author}
  {\bibfnamefont {T.}~\bibnamefont {Korn}}, \bibinfo {author} {\bibfnamefont
  {N.}~\bibnamefont {Paradiso}}, \bibinfo {author} {\bibfnamefont {M.~M.}\
  \bibnamefont {Glazov}},\ and\ \bibinfo {author} {\bibfnamefont
  {A.}~\bibnamefont {Chernikov}},\ }\href
  {https://doi.org/10.1103/PhysRevLett.120.207401} {\bibfield  {journal}
  {\bibinfo  {journal} {Phys. Rev. Lett.}\ }\textbf {\bibinfo {volume} {120}},\
  \bibinfo {pages} {207401} (\bibinfo {year} {2018})}\BibitemShut {NoStop}%
\bibitem [{\citenamefont {Perea-Caus{\'{\i}}n}\ \emph
  {et~al.}(2019)\citenamefont {Perea-Caus{\'{\i}}n}, \citenamefont {Brem},
  \citenamefont {Rosati}, \citenamefont {Jago}, \citenamefont {Kulig},
  \citenamefont {Ziegler}, \citenamefont {Zipfel}, \citenamefont {Chernikov},\
  and\ \citenamefont {Malic}}]{PereaCausin2019}%
  \BibitemOpen
  \bibfield  {author} {\bibinfo {author} {\bibfnamefont {R.}~\bibnamefont
  {Perea-Caus{\'{\i}}n}}, \bibinfo {author} {\bibfnamefont {S.}~\bibnamefont
  {Brem}}, \bibinfo {author} {\bibfnamefont {R.}~\bibnamefont {Rosati}},
  \bibinfo {author} {\bibfnamefont {R.}~\bibnamefont {Jago}}, \bibinfo {author}
  {\bibfnamefont {M.}~\bibnamefont {Kulig}}, \bibinfo {author} {\bibfnamefont
  {J.~D.}\ \bibnamefont {Ziegler}}, \bibinfo {author} {\bibfnamefont
  {J.}~\bibnamefont {Zipfel}}, \bibinfo {author} {\bibfnamefont
  {A.}~\bibnamefont {Chernikov}},\ and\ \bibinfo {author} {\bibfnamefont
  {E.}~\bibnamefont {Malic}},\ }\href
  {https://doi.org/10.1021/acs.nanolett.9b02948} {\bibfield  {journal}
  {\bibinfo  {journal} {Nano Letters}\ }\textbf {\bibinfo {volume} {19}},\
  \bibinfo {pages} {7317} (\bibinfo {year} {2019})}\BibitemShut {NoStop}%
\bibitem [{\citenamefont {Tosi}\ \emph {et~al.}(2012)\citenamefont {Tosi},
  \citenamefont {Christmann}, \citenamefont {Berloff}, \citenamefont {Tsotsis},
  \citenamefont {Gao}, \citenamefont {Hatzopoulos}, \citenamefont {Savvidis},\
  and\ \citenamefont {Baumberg}}]{Tosi2012a}%
  \BibitemOpen
  \bibfield  {author} {\bibinfo {author} {\bibfnamefont {G.}~\bibnamefont
  {Tosi}}, \bibinfo {author} {\bibfnamefont {G.}~\bibnamefont {Christmann}},
  \bibinfo {author} {\bibfnamefont {N.~G.}\ \bibnamefont {Berloff}}, \bibinfo
  {author} {\bibfnamefont {P.}~\bibnamefont {Tsotsis}}, \bibinfo {author}
  {\bibfnamefont {T.}~\bibnamefont {Gao}}, \bibinfo {author} {\bibfnamefont
  {Z.}~\bibnamefont {Hatzopoulos}}, \bibinfo {author} {\bibfnamefont {P.~G.}\
  \bibnamefont {Savvidis}},\ and\ \bibinfo {author} {\bibfnamefont {J.~J.}\
  \bibnamefont {Baumberg}},\ }\href {https://doi.org/10.1038/nphys2182}
  {\bibfield  {journal} {\bibinfo  {journal} {Nat. Phys.}\ }\textbf {\bibinfo
  {volume} {8}},\ \bibinfo {pages} {190} (\bibinfo {year} {2012})}\BibitemShut
  {NoStop}%
\bibitem [{\citenamefont {Dreismann}\ \emph {et~al.}(2014)\citenamefont
  {Dreismann}, \citenamefont {Cristofolini}, \citenamefont {Balili},
  \citenamefont {Christmann}, \citenamefont {Pinsker}, \citenamefont {Berloff},
  \citenamefont {Hatzopoulos}, \citenamefont {Savvidis},\ and\ \citenamefont
  {Baumberg}}]{Dreismann2014}%
  \BibitemOpen
  \bibfield  {author} {\bibinfo {author} {\bibfnamefont {A.}~\bibnamefont
  {Dreismann}}, \bibinfo {author} {\bibfnamefont {P.}~\bibnamefont
  {Cristofolini}}, \bibinfo {author} {\bibfnamefont {R.}~\bibnamefont
  {Balili}}, \bibinfo {author} {\bibfnamefont {G.}~\bibnamefont {Christmann}},
  \bibinfo {author} {\bibfnamefont {F.}~\bibnamefont {Pinsker}}, \bibinfo
  {author} {\bibfnamefont {N.~G.}\ \bibnamefont {Berloff}}, \bibinfo {author}
  {\bibfnamefont {Z.}~\bibnamefont {Hatzopoulos}}, \bibinfo {author}
  {\bibfnamefont {P.~G.}\ \bibnamefont {Savvidis}},\ and\ \bibinfo {author}
  {\bibfnamefont {J.~J.}\ \bibnamefont {Baumberg}},\ }\href
  {https://doi.org/10.1073/pnas.1401988111} {\bibfield  {journal} {\bibinfo
  {journal} {Proc. Natl. Acad. Sci. U.S.A.}\ }\textbf {\bibinfo {volume}
  {111}},\ \bibinfo {pages} {8770} (\bibinfo {year} {2014})}\BibitemShut
  {NoStop}%
\bibitem [{\citenamefont {Savvidis}\ \emph {et~al.}(2000)\citenamefont
  {Savvidis}, \citenamefont {Baumberg}, \citenamefont {Stevenson},
  \citenamefont {Skolnick}, \citenamefont {Whittaker},\ and\ \citenamefont
  {Roberts}}]{Savvidis2000a}%
  \BibitemOpen
  \bibfield  {author} {\bibinfo {author} {\bibfnamefont {P.~G.}\ \bibnamefont
  {Savvidis}}, \bibinfo {author} {\bibfnamefont {J.~J.}\ \bibnamefont
  {Baumberg}}, \bibinfo {author} {\bibfnamefont {R.~M.}\ \bibnamefont
  {Stevenson}}, \bibinfo {author} {\bibfnamefont {M.~S.}\ \bibnamefont
  {Skolnick}}, \bibinfo {author} {\bibfnamefont {D.~M.}\ \bibnamefont
  {Whittaker}},\ and\ \bibinfo {author} {\bibfnamefont {J.~S.}\ \bibnamefont
  {Roberts}},\ }\href {https://doi.org/10.1103/PhysRevLett.84.1547} {\bibfield
  {journal} {\bibinfo  {journal} {Phys. Rev. Lett.}\ }\textbf {\bibinfo
  {volume} {84}},\ \bibinfo {pages} {1547} (\bibinfo {year}
  {2000})}\BibitemShut {NoStop}%
\bibitem [{\citenamefont {Porras}\ \emph {et~al.}(2002)\citenamefont {Porras},
  \citenamefont {Ciuti}, \citenamefont {Baumberg},\ and\ \citenamefont
  {Tejedor}}]{Porras2002}%
  \BibitemOpen
  \bibfield  {author} {\bibinfo {author} {\bibfnamefont {D.}~\bibnamefont
  {Porras}}, \bibinfo {author} {\bibfnamefont {C.}~\bibnamefont {Ciuti}},
  \bibinfo {author} {\bibfnamefont {J.~J.}\ \bibnamefont {Baumberg}},\ and\
  \bibinfo {author} {\bibfnamefont {C.}~\bibnamefont {Tejedor}},\ }\href
  {https://doi.org/10.1103/PhysRevB.66.085304} {\bibfield  {journal} {\bibinfo
  {journal} {Phys. Rev. B}\ }\textbf {\bibinfo {volume} {66}},\ \bibinfo
  {pages} {085304} (\bibinfo {year} {2002})}\BibitemShut {NoStop}%
\bibitem [{\citenamefont {Rozas}\ \emph {et~al.}(2019)\citenamefont {Rozas},
  \citenamefont {Mart{\'{\i}}n}, \citenamefont {Tejedor}, \citenamefont
  {Vi{\~{n}}a}, \citenamefont {Deligeorgis}, \citenamefont {Hatzopoulos},\ and\
  \citenamefont {Savvidis}}]{Rozas2019}%
  \BibitemOpen
  \bibfield  {author} {\bibinfo {author} {\bibfnamefont {E.}~\bibnamefont
  {Rozas}}, \bibinfo {author} {\bibfnamefont {M.~D.}\ \bibnamefont
  {Mart{\'{\i}}n}}, \bibinfo {author} {\bibfnamefont {C.}~\bibnamefont
  {Tejedor}}, \bibinfo {author} {\bibfnamefont {L.}~\bibnamefont {Vi{\~{n}}a}},
  \bibinfo {author} {\bibfnamefont {G.}~\bibnamefont {Deligeorgis}}, \bibinfo
  {author} {\bibfnamefont {Z.}~\bibnamefont {Hatzopoulos}},\ and\ \bibinfo
  {author} {\bibfnamefont {P.~G.}\ \bibnamefont {Savvidis}},\ }\href
  {https://doi.org/10.1002/pssb.201800519} {\bibfield  {journal} {\bibinfo
  {journal} {Phys. Status Solidi B}\ }\textbf {\bibinfo {volume} {256}},\
  \bibinfo {pages} {1800519} (\bibinfo {year} {2019})}\BibitemShut {NoStop}%
\bibitem [{\citenamefont {Sun}\ \emph {et~al.}(2017{\natexlab{b}})\citenamefont
  {Sun}, \citenamefont {Yoon}, \citenamefont {Steger}, \citenamefont {Liu},
  \citenamefont {Pfeiffer}, \citenamefont {West}, \citenamefont {Snoke},\ and\
  \citenamefont {Nelson}}]{Sun2017a}%
  \BibitemOpen
  \bibfield  {author} {\bibinfo {author} {\bibfnamefont {Y.}~\bibnamefont
  {Sun}}, \bibinfo {author} {\bibfnamefont {Y.}~\bibnamefont {Yoon}}, \bibinfo
  {author} {\bibfnamefont {M.}~\bibnamefont {Steger}}, \bibinfo {author}
  {\bibfnamefont {G.}~\bibnamefont {Liu}}, \bibinfo {author} {\bibfnamefont
  {L.~N.}\ \bibnamefont {Pfeiffer}}, \bibinfo {author} {\bibfnamefont
  {K.}~\bibnamefont {West}}, \bibinfo {author} {\bibfnamefont {D.~W.}\
  \bibnamefont {Snoke}},\ and\ \bibinfo {author} {\bibfnamefont {K.~A.}\
  \bibnamefont {Nelson}},\ }\href {https://doi.org/10.1038/nphys4148}
  {\bibfield  {journal} {\bibinfo  {journal} {Nat. Phys.}\ }\textbf {\bibinfo
  {volume} {13}},\ \bibinfo {pages} {870} (\bibinfo {year}
  {2017}{\natexlab{b}})}\BibitemShut {NoStop}%
\bibitem [{\citenamefont {Cristofolini}\ \emph {et~al.}(2013)\citenamefont
  {Cristofolini}, \citenamefont {Dreismann}, \citenamefont {Christmann},
  \citenamefont {Franchetti}, \citenamefont {Berloff}, \citenamefont {Tsotsis},
  \citenamefont {Hatzopoulos}, \citenamefont {Savvidis},\ and\ \citenamefont
  {Baumberg}}]{Cristofolini2013}%
  \BibitemOpen
  \bibfield  {author} {\bibinfo {author} {\bibfnamefont {P.}~\bibnamefont
  {Cristofolini}}, \bibinfo {author} {\bibfnamefont {A.}~\bibnamefont
  {Dreismann}}, \bibinfo {author} {\bibfnamefont {G.}~\bibnamefont
  {Christmann}}, \bibinfo {author} {\bibfnamefont {G.}~\bibnamefont
  {Franchetti}}, \bibinfo {author} {\bibfnamefont {N.~G.}\ \bibnamefont
  {Berloff}}, \bibinfo {author} {\bibfnamefont {P.}~\bibnamefont {Tsotsis}},
  \bibinfo {author} {\bibfnamefont {Z.}~\bibnamefont {Hatzopoulos}}, \bibinfo
  {author} {\bibfnamefont {P.~G.}\ \bibnamefont {Savvidis}},\ and\ \bibinfo
  {author} {\bibfnamefont {J.~J.}\ \bibnamefont {Baumberg}},\ }\href
  {https://doi.org/10.1103/PhysRevLett.110.186403} {\bibfield  {journal}
  {\bibinfo  {journal} {Phys. Rev. Lett.}\ }\textbf {\bibinfo {volume} {110}},\
  \bibinfo {pages} {186403} (\bibinfo {year} {2013})}\BibitemShut {NoStop}%
\bibitem [{\citenamefont {Berloff}\ \emph {et~al.}(2017)\citenamefont
  {Berloff}, \citenamefont {Silva}, \citenamefont {Kalinin}, \citenamefont
  {Askitopoulos}, \citenamefont {T\"{o}pfer}, \citenamefont {Cilibrizzi},
  \citenamefont {Langbein},\ and\ \citenamefont {Lagoudakis}}]{Berloff2017}%
  \BibitemOpen
  \bibfield  {author} {\bibinfo {author} {\bibfnamefont {N.~G.}\ \bibnamefont
  {Berloff}}, \bibinfo {author} {\bibfnamefont {M.}~\bibnamefont {Silva}},
  \bibinfo {author} {\bibfnamefont {K.}~\bibnamefont {Kalinin}}, \bibinfo
  {author} {\bibfnamefont {A.}~\bibnamefont {Askitopoulos}}, \bibinfo {author}
  {\bibfnamefont {J.~D.}\ \bibnamefont {T\"{o}pfer}}, \bibinfo {author}
  {\bibfnamefont {P.}~\bibnamefont {Cilibrizzi}}, \bibinfo {author}
  {\bibfnamefont {W.}~\bibnamefont {Langbein}},\ and\ \bibinfo {author}
  {\bibfnamefont {P.~G.}\ \bibnamefont {Lagoudakis}},\ }\href
  {https://doi.org/10.1038/nmat4971} {\bibfield  {journal} {\bibinfo  {journal}
  {Nat. Mater.}\ }\textbf {\bibinfo {volume} {16}},\ \bibinfo {pages} {1120}
  (\bibinfo {year} {2017})}\BibitemShut {NoStop}%
\bibitem [{\citenamefont {Sun}\ \emph {et~al.}(2018)\citenamefont {Sun},
  \citenamefont {Yoon}, \citenamefont {Khan}, \citenamefont {Ge}, \citenamefont
  {Steger}, \citenamefont {Pfeiffer}, \citenamefont {West}, \citenamefont
  {T\"{u}reci}, \citenamefont {Snoke},\ and\ \citenamefont {Nelson}}]{Sun2018}%
  \BibitemOpen
  \bibfield  {author} {\bibinfo {author} {\bibfnamefont {Y.}~\bibnamefont
  {Sun}}, \bibinfo {author} {\bibfnamefont {Y.}~\bibnamefont {Yoon}}, \bibinfo
  {author} {\bibfnamefont {S.}~\bibnamefont {Khan}}, \bibinfo {author}
  {\bibfnamefont {L.}~\bibnamefont {Ge}}, \bibinfo {author} {\bibfnamefont
  {M.}~\bibnamefont {Steger}}, \bibinfo {author} {\bibfnamefont {L.~N.}\
  \bibnamefont {Pfeiffer}}, \bibinfo {author} {\bibfnamefont {K.}~\bibnamefont
  {West}}, \bibinfo {author} {\bibfnamefont {H.~E.}\ \bibnamefont
  {T\"{u}reci}}, \bibinfo {author} {\bibfnamefont {D.~W.}\ \bibnamefont
  {Snoke}},\ and\ \bibinfo {author} {\bibfnamefont {K.~A.}\ \bibnamefont
  {Nelson}},\ }\href {https://doi.org/10.1103/PhysRevB.97.045303} {\bibfield
  {journal} {\bibinfo  {journal} {Phys. Rev. B}\ }\textbf {\bibinfo {volume}
  {97}},\ \bibinfo {pages} {045303} (\bibinfo {year} {2018})}\BibitemShut
  {NoStop}%
\end{thebibliography}
\end{document}